\documentclass[11pt]{amsart}
\usepackage{kbordermatrix}
\usepackage{graphicx}
\usepackage{wrapfig}
\usepackage{amssymb}
\usepackage{algorithm}
\usepackage{mathtools}
\usepackage{textcomp}
\usepackage[shortlabels]{enumitem}
\usepackage{times}
\usepackage[top=20mm, bottom=20mm, left=25mm, right=25mm]{geometry}
\usepackage{color}
\usepackage{bm}
\usepackage{algorithmic}
\usepackage{subcaption}
\usepackage{booktabs}

\newlength{\topFigureVerticalSpace}
\newlength{\bottomFigureVerticalSpace}
\newcommand{\tfvspace}{\vspace*{\topFigureVerticalSpace}}
\newcommand{\bfvspace}{\vspace*{\bottomFigureVerticalSpace}}

\newcommand{\vs}{\vspace*{2mm}}
\newcommand{\minusvs}{\vspace*{-2mm}}

\newcommand{\mtx}[1]{\mathnormal{#1}}

\newcommand{\defeq}{\vcentcolon=}
\newcommand{\si}{(i)} 
\newcommand{\simo}{(i-1)} 

\theoremstyle{definition}
\newtheorem{remark}{Remark}
\newtheorem{exercise}{Exercise}

\newcommand{\randUTV}{\texttt{randUTV}}

\usepackage{ifthen}
\usepackage{colortbl}
\usepackage{array}
\usepackage{color}
\usepackage{soul}
\usepackage{amssymb}

\newboolean{IsWide}
\setboolean{IsWide}{true}

\newcommand{\FlaTwoByTwo}[4]{
\left(
\begin{array}{c I c}
#1 & #2 \\ \whline
#3 & #4
\end{array}
\right)
}

\newcommand{\FlaTwoByTwoSingleLine}[4]{
\left(
\begin{array}{c | c}
#1 & #2 \\ \hline
#3 & #4
\end{array}
\right)
}

\newcommand{\FlaTwoByOneSingleLine}[2]{
\left(
\begin{array}{c}
#1 \\ \hline
#2
\end{array}
\right)
}

\newcommand{\FlaOneByTwo}[2]{
\left(
\begin{array}{c I c}
#1 & #2
\end{array}
\right)
}

\newcommand{\FlaOneByTwoSingleLine}[2]{
\left(
\begin{array}{c | c}
#1 & #2
\end{array}
\right)
}

\newcommand{\FlaThreeByThreeTL}[9]{
\left(
\begin{array}{c | c I c}
#1 & #2 & #3 \\ \hline
#4 & #5 & #6 \\ \whline
#7 & #8 & #9
\end{array}
\right)
}

\newcommand{\FlaThreeByThreeBR}[9]{
\left(
\begin{array}{c I c | c}
#1 & #2 & #3 \\ \whline
#4 & #5 & #6 \\ \hline
#7 & #8 & #9
\end{array}
\right)
}

\newcommand{\FlaOneByThreeR}[3]{
\left(
\begin{array}{c I c | c}
#1 & #2 & #3
\end{array}
\right)
}

\newcommand{\FlaOneByThreeL}[3]{
\left(
\begin{array}{c | c I c}
#1 & #2 & #3
\end{array}
\right)
}

\newcommand{\FlaPartition}[2]{
\ifthenelse{\boolean{IsWide}}{{\bf partition } \hspace{-1em} #1 \hspace{-1em} #2}
{{\bf partition } \+ \\ #1 \+ \\ #2 \- \-}
}

\newcommand{\FlaRepartition}[2]{
\ifthenelse{\boolean{IsWide}}{{\bf repartition } \hspace{-1em} #1 \hspace{-1em} #2}
{{\bf repartition } \+ \\ #1 \+ \\ #2 \- \-}
}

\newcommand{\FlaStartCompute}{%
\setlength{\unitlength}{1.6in}%
\begin{picture}(3,0.01)
\put(0,0){\line(1,0){3}}
\put(0,0.01){\line(1,0){3}}
\end{picture}%
}

\newcommand{\FlaEndCompute}{%
\noindent%
\setlength{\unitlength}{1.6in}%
\begin{picture}(3,0.01)
\put(0,0){\line(1,0){3}}
\put(0,0.01){\line(1,0){3}}
\end{picture}%
}

\newcommand{\operation}{ [ D, E, F, \ldots ] \becomes {\rm op}( A, B, C, D, \ldots ) }
\newcommand{\routinename}{ [ D, E, F, \ldots ] \becomes {\rm op}( A, B, C, D, \ldots ) }
\newcommand{\routinecost}{ X }
\newcommand{\precondition}{ Q }
\newcommand{\postcondition}{ R }
\newcommand{\invariant}{ P }
\newcommand{\costinv}{ \  }

\newcommand{\guard}{ R }
\newcommand{\partitionings}{
\begin{minipage}{3in}
$ S_I $
\end{minipage}
}
\newcommand{\initialize}{}

\newcommand{\partitionsizes}{ \hspace{ 3.25in} }

\newcommand{\blocksize}{1}

\newcommand{\repartitionings}{
\begin{minipage}[t]{3in}
\ \\
\ \\
\ \\
\end{minipage}
}

\newcommand{\repartitionsizes}{ \hspace{ 3.25in} }
\newcommand{\moveboundaries}{
\begin{minipage}[t]{3in}
\ \\
\ \\
\ \\
\end{minipage}
}
\newcommand{\beforeupdate}{
$ \QBefore $
}
\newcommand{\afterupdate}{
$ \QAfter $
}
\newcommand{\update}{%
\begin{minipage}[t]{4in}
$ S_U $
\end{minipage}
}

\newcommand{\resetsteps}{
\renewcommand{\blocksize}{1}
\renewcommand{\operation}{ [ D, E, F, \ldots ] \becomes {\rm op}( A, B, C, D, \ldots ) }
\renewcommand{\routinename}{ [ D, E, F, \ldots ] \becomes {\rm op}( A, B, C, D, \ldots ) }
\renewcommand{\routinecost}{ 0 }
\renewcommand{\precondition}{ \PPre }
\renewcommand{\postcondition}{ \PPost }
\renewcommand{\invariant}{ \PInv }
\renewcommand{\costinv}{ \  }
\renewcommand{\guard}{ G }
\renewcommand{\partitionings}{ %
\begin{minipage}[t]{3in}
\ \\
\end{minipage}
}
\renewcommand{\partitionsizes}{ \hspace{ 3.25in} }
\renewcommand{\repartitionings}{%
\begin{minipage}[t]{3in}
\ \\
\end{minipage}
}
\renewcommand{\repartitionsizes}{ \hspace{ 3.25in} }
\renewcommand{\moveboundaries}{%
\begin{minipage}[t]{3in}
\ \\
\end{minipage}
}
\renewcommand{\beforeupdate}{
\QBefore
}
\renewcommand{\afterupdate}{
\QAfter
}
\renewcommand{\update}{
$ S_U $
}
}

\newcommand{\WSguard}{
$ \guard $
}

\newcommand{\WSpartition}{%
\begin{minipage}[t]{3.0in}%
\begin{tabbing}
ind \= ind \= \kill
{\bf \color{blue} Partition}
\partitionings \+ \\
{\bf \color{blue} where } \hspace*{-2ex} \partitionsizes
\end{tabbing}
\end{minipage}
}

\newcommand{\WSinitialize}{
\initialize
}

\newcommand{\WSrepartition}{
\begin{minipage}[t]{3in}
\ifthenelse{ \equal{\blocksize}{1} }{}
{%
\ifthenelse{ \equal{\blocksize}{blank} }{~}
{{\bf \color{blue} Determine block size} $ \blocksize $} \\
}
{\bf \color{blue} Repartition}
\begin{tabbing}
in \= in \= \+ \kill
\repartitionings \+ \\
{\bf \color{blue} where } \hspace*{-2ex} \repartitionsizes
\end{tabbing}
\end{minipage}
}

\newcommand{\WSrepartitionNarrow}{
\begin{minipage}[t]{2.05in}
\ifthenelse{ \equal{\blocksize}{1} }{\phantom{Determine}~}
{%
\ifthenelse{ \equal{\blocksize}{blank} }{}
{{\bf \color{blue} Determine block size} $ \blocksize $} \\
}
{\bf \color{blue} Repartition}
\begin{tabbing}
i \= i \= \+ \kill
\repartitionings 
\+ \\
{\bf \color{blue} where }
\begin{minipage}[t]{1.5in}
\repartitionsizes
\end{minipage}
\end{tabbing}
\end{minipage}
}

\newcommand{\WSmoveboundary}{%
\begin{minipage}[t]{4in}%
{\bf \color{blue} Continue with}
\begin{tabbing}
ind \= \+ \kill
\moveboundaries
\end{tabbing}
\end{minipage}
}

\newcommand{\WSupdate}{
\update
}

\newcommand{\FlaAlgorithmWithInit}{
\begin{center}
\begin{tabular}{| p{0.98\textwidth} |} \hline
{\bf \color{blue} Algorithm:} $\routinename$
\\ \whline
\WSinitialize \\
{\WSpartition} \\[0.3in]
{\bf \color{blue} while} \WSguard { \bf \color{blue} do} \\
\ \hspace{0.15in} \WSrepartition \\
\color{red} {\hspace{0.0in} \FlaStartCompute} \\
{\hspace{0.0in} \WSupdate} \\
\color{red} {\hspace{0.0in} \FlaEndCompute} \\
{\ \hspace{0.15in} \WSmoveboundary} \\
{{\bf \color{blue} endwhile}} \\ \hline
\end{tabular}
\end{center}
}

\newcommand{ \PPre }{ P_{\it pre} }
\newcommand{ \PPost }{ P_{\it post} }
\newcommand{ \PInv }{ P_{\it inv} }

\newcommand{ \QBefore }{ P_{\it before} }
\newcommand{ \QAfter }{ P_{\it after} }

\newcommand{\Answer}[1]{%
\ifthenelse{\boolean{ShowAnswers}}{%
{\color{blue}%
\vspace{0.05in}%
\noindent%
{\bf Answer:} #1}}
{}%
}

\newcommand{\PartAnswer}[1]{%
\ifthenelse{\boolean{ShowAnswers}}{%
{\color{blue} #1}}%
{}
}

\newcommand{\QuestionAnswer}[2]{%
\ifthenelse{\boolean{ShowAnswers}}{%
{\color{blue} #2}}%
{\color{black} #1}
}

\newcommand{\ShowAnswer}[1]{
\ifthenelse{\boolean{ShowAnswers}}{
{\color{blue} #1}}
{\phantom{#1}}
}

{
\vspace{0.1in}
\end{minipage}
\\ \whline
\end{tabular}
}

\newenvironment{unboxedexercise}{
\addtocounter{homeworkcounter}{1}%
\noindent%
\begin{exercise}
\rm 
}
{
\end{exercise}%
}

\newcolumntype{I}{!{\vrule width 1.5pt}}
\newlength\savedwidth
\newcommand\whline{\noalign{\global\savedwidth\arrayrulewidth
                            \global\arrayrulewidth 1.5pt}%
           \hline
           \noalign{\global\arrayrulewidth\savedwidth}}

\newcommand{\NoShow}[1]{}

\newcommand{\becomes}{:=}

\begin{document}

\begin{center}
\textbf{\large{Computing rank-revealing factorizations of matrices
               stored out-of-core}}

\vspace{5mm}

\textit{
  N.~Heavner~\footnotemark[1],
  P.G.~Martinsson~\footnotemark[2],
  G.~Quintana-Ort\'{\i}~\footnotemark[3]
}

\vspace{5mm}

\begin{minipage}{140mm}
\textbf{Abstract:}
This paper describes efficient algorithms for computing rank-revealing
factorizations of matrices that are too large to fit in RAM, and must
instead be stored on slow external memory devices such as solid-state
or spinning disk hard drives (out-of-core or out-of-memory).
Traditional algorithms for computing rank revealing
factorizations, such as the column pivoted QR factorization, or techniques
for computing a full singular value decomposition of a matrix, are very
communication intensive. They are naturally expressed as a sequence of
matrix-vector operations, which become prohibitively expensive when data
is not available in main memory. Randomization allows these methods to
be reformulated so that large contiguous blocks of the matrix can be
processed in bulk.
The paper describes two distinct methods.
The first is a blocked version of column pivoted Householder QR, organized
as a ``left-looking'' method to minimize the number of write operations
(which are more expensive than read operations on a spinning disk drive).
The second method results in a so called UTV factorization which expresses
a matrix $\mtx{A}$ as $\mtx{A} = \mtx{U}\mtx{T}\mtx{V}^{*}$ where
$\mtx{U}$ and $\mtx{V}$ are unitary, and $\mtx{T}$ is triangular.
This method is organized as an algorithm-by-blocks, in which floating
point operations overlap read and write operations. 
The second method incorporates power iterations, and is exceptionally good at revealing the numerical rank;
it can often be used as a substitute for a full singular value decomposition.
Numerical experiments demonstrate that the new algorithms are almost
as fast when processing data stored on a hard drive as traditional
algorithms are for data stored in main memory. To be precise, the
computational time for fully factorizing an $n\times n$ matrix
scales as $cn^{3}$, with a scaling constant $c$ that is only marginally
larger when the matrix is stored out of core.
\end{minipage}

\end{center}

\vspace{5mm}


\footnotetext[1]{
  Department of Applied Mathematics, 
  University of Colorado at Boulder, 
  526 UCB, Boulder, CO 80309-0526, USA
}

\footnotetext[2]{
  Department of Mathematics,
  University of Texas at Austin,
  Stop C1200,
  Austin, TX 78712-1202, USA
}

\footnotetext[3]{
  Departamento de Ingenier\'{\i}a y Ciencia de Computadores, 
  Universitat Jaume I, 
  12.071 Castell\'on, Spain
}

\section{Introduction}

\subsection{Problem formulation}
We consider the task of computing a rank-revealing factorization of a matrix
that is so large that it must be stored on an external memory device such as a spinning 
or solid-state disk drive. In this case, the matrix is said to be stored \textit{out-of-core}.
Specifically, given an $m \times n$ matrix $\mtx{A}$, we seek a factorization of the form
\begin{equation}
\label{eq:basic}
  \begin{array}{ccccccccccc}
    \mtx{A} &=& \mtx{U} & \mtx{R} & \mtx{V}^{*},\\
    m\times n && m\times m & m\times n & n\times n
  \end{array}
\end{equation}
where $\mtx{U}$ and $\mtx{V}$ are unitary matrices, and where $\mtx{R}$ is upper triangular.
We use the term ``rank-revealing'' in an informal way, as meaning simply that for any 
$k$ with $1 \le k \le \min(m,n)$, the truncation of (\ref{eq:basic}) to the $k$ dominant
terms provides a rank-$k$ approximation to $\mtx{A}$ that is almost as good as the
theoretically optimal one. To be precise, with $\mtx{A}_k = \mtx{U}(:,1:k) \mtx{R}(1:k,:) \mtx{V}^*$,
we ask that 
$$
\| \mtx{A} - \mtx{A}_k \| \approx \inf\{\|\mtx{A} - \mtx{B}\|\,\colon\,\mbox{where }\mtx{B}\mbox{ has rank }k\}.
$$
Problems which make use of such a factorization include
solving ill-conditioned linear systems, rank-deficient and
total least squares problems~\cite{lawson1995solving,golub1980analysis,van1991total,lawson1995solving},
subset selection~\cite{miller2002subset}, and
matrix approximation~\cite{eckart1936approximation,mahoney2011randomized},
among others.

\subsection{Prior work}
There are several well-established options for computing a
rank-revealing factorization.
If a full factorization is required, the singular value decomposition (SVD)
and the QR decomposition with column pivoting (CPQR) are two
popular options.
The SVD provides theoretically optimal low rank approximations for many choices of matrix norms,
but can be prohibitively expensive to compute.
The CPQR usually reveals rank reasonably well
(see, \textit{e.g.}~\cite{kahan1966numerical} for a notable exception)
and requires much less computational work than the SVD.
Neither factorization is easily implemented
when the matrix is stored out-of-core, however.
The traditional algorithms for computing each require
many slow matrix-vector operations,
which in turn necessitate
many read operations (data brought from an external device to main memory)
and write operations (data sent from main memory to an external device).
See, \textit{e.g.} ~\cite{trefethen1997numerical,golub,klema1980singular}
for information on standard algorithms for computing the SVD, and
\cite{golub1965numerical,stewart1998matrix,quintana1998blas}
for information on the most common CPQR algorithm and its
restriction to matrix-vector operations.

Some software efforts in computing factorizations
of dense matrices stored in external devices include
POOCLAPACK~\cite{gunter2005parallel,reiley1999pooclapack,gunter2001parallel},
the SOLAR~\cite{toledo1996design} library,
a runtime to execute algorithms-by-blocks
when the matrix is stored in disk~\cite{quintana2012oocruntime},
as well as an out-of-core extension~\cite{d2000design}
to ScaLAPACK~\cite{blackford1997scalapack}
that does not appear in the current version of the library.
Even for these libraries, factorization routines
are usually limited to Cholesky, unpivoted QR, and
LU decompositions.
A truncated SVD algorithm that is effective for out-of-core matrices
was introduced in~\cite{halko2011algorithm}.
This technique, however, only yields a \textit{partial} factorization, and
it requires that an upper bound for the target rank $k$ is known in advance.
Recent efforts have been able to compute the SVD of large dense matrices
stored in an external device.
Demchik \textit{et al.}~\cite{demchik:2019:svd}
employed randomization to compute the SVD,
but no performance results for dense matrices were shown.
Kabir \textit{et al.}~\cite{kabir:2017:svd}
used a two-stage approach to compute the SVD
that reduces the dense matrix to band form in a first stage,
which allowed to factorize
matrices of up to dimension $100k \times 100k$.
To our knowledge,
there are no currently available and widely used software
for computing a full rank-revealing factorization out-of-core.

\subsection{Contributions of present work}
This article describes two different algorithms for computing
rank-revealing factorizations of matrices stored out-of-core.
The first, \texttt{HQRRP\_left},
is based on the \texttt{HQRRP} algorithm~\cite{martinsson2017householder},
which uses randomization techniques
to build a fully blocked column pivoted QR factorization.
\texttt{HQRRP} relies largely on matrix-matrix operations,
reducing the number of reads and writes that must occur
between the external device and main memory (RAM).
\texttt{HQRRP\_left} further reduces the number of write operations
required by redesigning \texttt{HQRRP} as a left-looking algorithm.
Numerical experiments reveal that this reduction in the writing time
is critical to the algorithm's performance,
particularly when the data is stored on a spinning disk.
(The \texttt{HQRRP} method is closely related to the technique in \cite{2017_blockQR_ming_article}.)

The second contribution of this article is an out-of-core implementation
of the \randUTV{} algorithm~\cite{2019_martinsson_randUTV_ACMTOMS},
which uses randomization to build a rank-revealing UTV factorization
(see, \textit{e.g.}~\cite{1994_stewart_UTV,stewart1992updating}
for a good introduction to this factorization).
The out-of-core implementation, \texttt{randUTV\_AB},
modifies \randUTV{} in such a way as to achieve
overlap of communication and floating point operations.
The result, demonstrated with numerical experiments,
is that \texttt{randUTV\_AB} suffers only a small extra cost
by storing the matrix in an external device
with respect to the same factorization of a matrix stored in RAM.

The new techniques presented enable processing of very large matrices
at modest cost. To illustrate, at the time of writing in 2019, a personal 
desktop with a 2TB high speed SSD (like the one in the machine ``ut'' in 
Section \ref{sec:expsetup}) can be had for a couple of thousands of dollars, 
whereas a workstation with an equivalent amount of RAM would be an order
of magnitude more expensive. Since solid state storage technology is rapidly 
getting both cheaper and faster, we expect to see demand for out-of-core 
algorithms to continue to increase.

\subsection{Assessing the quality of a rank-revealing factorization}

The term ``rank-revealing factorization'' has been used with slightly
different meanings in the literature~\cite{chandrasekaran1994rank,stewart1992updating,chan1987rank,chan1992some};
our requirement that the approximation error $\|\mtx{A} - \mtx{A}_k\|$ is close to optimal is
invariably a part of the requirement, but in more theoretical papers the statement is often
made precise by bounding how rapidly the discrepancy is allowed to increase as the matrix
dimensions tend to infinity. 


The focus of the current paper is how to efficiently implement algorithms that 
have already been published. The precision at which they reveal rank has been
carefully studied: The randomized CPQR was analyzed in \cite{2017_blockQR_ming_article}, 
with additional numerical results presented in \cite{martinsson2017householder}.
The conclusion of these papers was that the randomized pivoting reveals the
numerical rank to roughly the same precision as classical pivoting.
The randomized UTV factorization was studied in \cite{2019_martinsson_randUTV_ACMTOMS},
with additional results reported in \cite{2019_heavner_diss}. The precision of
\randUTV{} depends on a tuning parameter $q$ that controls the number of power
iterations that are taken at each step. Higher $q$ results in better precision,
but a slower execution time. When no powering is done ($q=0$), \randUTV{} is
about as good at revealing the rank as CPQR, with \randUTV{} often having a 
slight edge. As $q$ is increased to one or two, the precision very rapidly
improves, and often gets to within striking distance of the optimal precision
of the SVD. Moreover, the diagonal entries of the middle factor $\mtx{T}$ often
provide excellent approximations to the true singular values of the matrix.


\subsection{Outline of paper}
In Section~\ref{sec:hqrrp}, we discuss the first out-of-core factorization,
a rank-revealing QR factorization \texttt{HQRRP}
which can be stopped early to yield a partial decomposition.
Section~\ref{sec:randutv} explores an out-of-core implementation of
\randUTV{}, an efficient algorithm for obtaining
a rank-revealing orthogonal matrix decomposition.
In Section~\ref{sec:num} we present numerical experiments
which demonstrate the performance of both algorithms.
Finally, Section~\ref{sec:conclusions}
contains the main conclusions of our work.

\section{Partial rank-revealing QR factorization}
\label{sec:hqrrp}

In this section we introduce an out-of-core implementation of
a rank-revealing column pivoted QR factorization.
In subsection~\ref{sec:hqrrp-overview}
and subsection~\ref{sec:building-p-and-q},
we review a fully blocked algorithm
for computing a column pivoted QR factorization, \texttt{HQRRP}.
In subsection~\ref{sec:hqrrp-ooc}, we discuss modifications of \texttt{HQRRP}
which enhance its efficiency when the matrix is stored out-of-core.

\subsection{Overview of \texttt{HQRRP}}
\label{sec:hqrrp-overview}

Consider an input matrix $\mtx{A} \in \mathbb{R}^{m \times n}$ with $m \ge n$.
\texttt{HQRRP}~\cite{martinsson2017householder} is a \textit{blocked}
algorithm for computing a column-pivoted QR factorization
\[
  \begin{array}{ccccc}
    \mtx{A} & = & \mtx{Q} & \mtx{R} & \mtx{P}^*, \\
    m \times n & & m \times m & m \times n & n \times n
  \end{array}
\]
where $\mtx{Q}$ is orthogonal,
$\mtx{R}$ is upper trapezoidal, and
$\mtx{P}$ is a permutation matrix.

The bulk of the algorithm's work is executed
in a loop with $\lceil n/b \rceil$ iterations,
where $1 \le b \le n$ is the \textit{block size} parameter.
For notational simplicity, for the remaining discussion
it is assumed that $n$ is a multiple of $b$ so that $n = b p$ for some $p \in \mathbb{N}$.
At the start of the \texttt{HQRRP} algorithm are the following initializations:
\[
  \mtx{R}^{(0)} = \mtx{A}, \quad
  \mtx{Q}^{(0)} = \mtx{I}, \quad
  \mtx{P}^{(0)} = \mtx{I}.
\]

During the $i$-th iteration, matrices $\mtx{Q}^{\si}$ and $\mtx{P}^{\si}$
are constructed such that
\[
  \mtx{R}^{\si} = (\mtx{Q}^{\si})^* \mtx{R}^{\simo} \mtx{P}^{\si},
\]
where $\mtx{R}^{\si}(:,1:ib)$ is upper trapezoidal.
After $ p $ steps, $\mtx{R}^{(p)}$ is upper trapezoidal, and the final
factorization can be written as
\begin{align*}
  \mtx{R} & = \mtx{R}^{( p )} \\
  \mtx{P} & = \mtx{P}^{(0)} \mtx{P}^{(1)} \cdots \mtx{P}^{( p )} \\
  \mtx{Q} & = \mtx{Q}^{(0)} \mtx{Q}^{(1)} \cdots \mtx{Q}^{( p )}.
\end{align*}

\subsection{Choosing the orthogonal matrices}
\label{sec:building-p-and-q}

At the $i$-th step of \texttt{HQRRP},
consider the partitioning of $\mtx{R}^{\si}$
\[
  \mtx{R}^{\si}
  \rightarrow
  \begin{pmatrix}
    \mtx{R}^{\si}_{11} & \mtx{R}^{\si}_{12} \\
    \mtx{R}^{\si}_{11} & \mtx{R}^{\si}_{22}
  \end{pmatrix},
\]
where $\mtx{R}^{\si}$ is $ib \times ib$.
The permutation matrix $\mtx{P}^{\si}$ is chosen to find $b$ columns
of $\mtx{R}^{\simo}_{22}$ with a large spanning volume,
in the sense that the spanning volume is close to maximal.
We can find such a selection by projecting the columns
of $\mtx{R}^{\simo}_{22}$ into a lower dimensional space and
cheaply finding $b$ good pivot columns there.
The steps for computing $\mtx{P}^{\si}$ are thus as follows:

\begin{enumerate}[1.]

\item
Draw a random matrix $\mtx{G}^{\si} \in \mathbb{R}^{b \times (m - ib)}$,
with i.i.d.~entries drawn from the standard normal distribution.

\item
Compute $\mtx{Y}^{\si} = \mtx{G}^{\si} \mtx{R}^{\simo}_{22}$.

\item
Compute $b$ steps of the traditional CPQR of $\mtx{Y}^{\si}$ to obtain
$\mtx{Y}^{\si} \mtx{P}_{22}^{\si} =
 \mtx{W}_{\text{trash}} \mtx{S}_{\text{trash}}$.

\item
Set
\[
  \mtx{P}^{\si} =
  \begin{pmatrix}
    \mtx{I} & 0 \\
    0 & \mtx{P}_{22}^{\si}
  \end{pmatrix}.
\]

\end{enumerate}

This method for selecting multiple pivot columns has shown itself
to be effective and reliable, consistently producing factorizations
that reveal the rank
as well as traditional CPQR~\cite{martinsson2017householder}.

\begin{remark}
There is an alternate ``downdating'' method
for computing $\mtx{Y}^{\si}$ during each step that reduces the
asymptotic flop count of the \texttt{HQRRP} algorithm
\cite{2017_blockQR_ming_article}.
With this technique,
\texttt{HQRRP} has the same asymptotic flop count
as the unpivoted QR factorization and the CPQR;
the reader may see~\cite{martinsson2017householder} for details.
However,
this downdating method will not be used
in this article's primary implementation
as the communication restrictions imposed by this downdating method
make the basic scheme discussed in this section more practical.
\end{remark}

Once $\mtx{P}^{\si}$ has been computed,
$\mtx{Q}^{\si}$ is built with well-established steps:

\begin{enumerate}[1.]

\item
Perform unpivoted QR factorization on
$\mtx{A}^{\simo}_{22} \mtx{P}_{22}(:,1:b)$
to obtain
\[
  \mtx{A}^{\simo}_{22} \mtx{P}_{22}
  =
  \mtx{Q}_{22}^{\si} \mtx{A}^{\si}_{22}(:,1:b).
\]

\item
Set
\[
  \mtx{Q}^{\si}
  =
  \begin{pmatrix}
    \mtx{I} & 0 \\
    0 & \mtx{Q}_{22}^{\si}
  \end{pmatrix}.
\]

\end{enumerate}

\subsection{Executing \texttt{HQRRP} out-of-core}
\label{sec:hqrrp-ooc}

When the matrix to be factorized is large enough
that it must be stored out-of-core,
communication (I/O operations) costs become a major concern.
While it is desirable to minimize all forms of communication,
writing to an external device is typically more expensive than reading from it.
%
In linear algebra,
in every row, column, or block iteration,
right-looking algorithms factorize the current row, column, or block and
then update the rest of the matrix,
which usually requires an overall cost of $\mathcal{O} (n^3)$ writes.
In contrast,
left-looking algorithms apply all the previous transformations
to the current row, column, or block and then factorize it,
which usually requires an overall cost of only $\mathcal{O} (n^2)$ writes.
When working on main memory,
performances are only slightly different,
but when working with matrices stored in an external device
with slow writes, left-looking algorithms deliver higher performances.

The original \texttt{HQRRP} was a classical right-looking algorithm,
and therefore it performed many write operations.
We have designed a left-looking variation of
the original \texttt{HQRRP} algorithm,
so that the number of write operations are much smaller.

\subsubsection{Left-looking algorithms}
\label{sec:left-looking}

Several standard matrix factorizations have been re-ordered
as left-looking algorithms for out-of-core computations,
largely with the goal of
reducing certain I/O operations~\cite{reiley1999efficient,
reiley1999pooclapack,toledo1996design,kurzak2010scheduling}.

As our new algorithm employs the QR factorization,
a high level description of the left-looking algorithm
for computing the QR factorization follows.
Let $\mtx{A} \in \mathbb{R}^{n \times n}$ and
$b$ be a block size $1 \le b \le n$;
for simplicity let $n = b p$, where $b,p \in \mathbb{N}$.
Then for $i=0, 1, \ldots, p-1,$
the algorithm for computing the QR factorization is described
by the following steps (indices start at zero):

\begin{enumerate}

\item
$\mtx{A}_{\text{col}} \leftarrow \mtx{A}(:, ib:(i+1)b-1)$.

\item
$\mtx{A}_{\text{col}} \leftarrow \mtx{Q}^* \mtx{A}_{\text{col}}$.

\item
Compute the unpivoted QR factorization of $\mtx{A}_{\text{col}}(ib:n-1,:)$
yielding $\mtx{Q}_i, \mtx{R}_i$.

\item
$\mtx{A}(0:ib-1,ib:(i+1)b-1) \leftarrow \mtx{A}_{\text{col}}(0:ib-1,:)$.

\item
$\mtx{A}(ib:n-1,ib:(i+1)b-1) \leftarrow \mtx{R}_i$.

\item
$\mtx{Q} \leftarrow
  \mtx{Q}
  \begin{pmatrix}
    \mtx{I} & 0 \\
    0 & \mtx{Q}_i
  \end{pmatrix}$.

\end{enumerate}

In this algorithm, $\mtx{Q}$ is initialized as the identity matrix $\mtx{I}$,
and after the algorithm is finished, $\mtx{A}$ is overwritten with
the upper triangular matrix $\mtx{R}$.
In practice, $\mtx{Q}$ is never formed, but instead the Householder vectors
are stored in the lower triangular portion of $\mtx{R}$ as it is computed.
This overview omits other computational details as well,
but clearly highlights the fact that
only the column block of $\mtx{A}(:,ib:(i+1)b-1)$ is updated in every iteration.
In contrast,
the traditional right-looking algorithm updates
the bottom-right block $\mtx{A}(ib:n-1,ib:n-1)$
during each iteration of the loop,
thus requiring much more writing to disk.

\subsubsection{\texttt{HQRRP} as a left-looking algorithm}
\label{sec:left-hqrrp}

\texttt{HQRRP} as a left-looking algorithm follows the pattern of the
unpivoted factorization discussed in Section~\ref{sec:left-looking},
with the added complication of choosing and applying
the permutation matrix $\mtx{P}$.
We select the permutations
using precisely the same procedure as in~\ref{sec:building-p-and-q}.
Observe that the projection procedure used in the selection of $\mtx{P}$
requires that the right-most columns of $\mtx{A}$ be updated according to
the rotations encoded in $\mtx{Q}$ during each iteration.
We must therefore choose whether to write out the updated columns of
$\mtx{A}$ to disk or repeat some of the arithmetic operations during each
iteration.
Since the main idea of a left-looking algorithm is to avoid writing to the
right-most columns of $\mtx{A}$ every iteration, we must accept the
repetition of operations as part of \texttt{HQRRP\_left} with the current
technologies.
Numerical experiments indicated that the I/O-saving benefits of
\texttt{HQRRP\_left} easily outweigh the costs of the extra flops as
compared to the original right-looking \texttt{HQRRP}.

\section{Full rank-revealing orthogonal factorization}
\label{sec:randutv}

In this section we introduce an efficient implementation of a rank-revealing
orthogonal decomposition for a matrix stored out-of-core.
In subsection~\ref{sec:randutv-overview},
we review an efficient algorithm, \randUTV{},
for building such a decomposition when the matrices are stored in main memory.
In subsection~\ref{sec:randutv-ooc},
we discuss some modifications of \randUTV{}
that optimize its efficiency in the out-of-core setting.

\subsection{Overview of \texttt{randUTV}}
\label{sec:randutv-overview}

Let $\mtx{A} \in \mathbb{R}^{m \times n}$ with $m \ge n $.
The \randUTV{} algorithm~\cite{2019_martinsson_randUTV_ACMTOMS}
builds a rank-revealing UTV factorization of $\mtx{A}$, that is,
a decomposition
\[
  \begin{array}{ccccc}
    \mtx{A} & = & \mtx{U} & \mtx{T} & \mtx{V}^* \\
    m \times n & & m \times m & m \times n & n \times n
  \end{array}
\]
such that $\mtx{U}$ and $\mtx{V}$ are orthogonal and
$\mtx{T}$ is upper triangular.
The \randUTV{} algorithm is blocked,
so it proceeds by choosing first a block size $b$
with $1 \le b \le n$ and performing its work inside a loop
with $\lceil n/b \rceil$ iterations.
For ease of notation, it is assumed that $n = b p$ for $b,p \in \mathbb{N}$.
At the beginning, we initialize
\[
  \mtx{T}^{(0)} = \mtx{A}, \quad
  \mtx{U}^{(0)} = \mtx{I}, \quad
  \mtx{V}^{(0)} = \mtx{I}.
\]

For the matrix $\mtx{T}^{\si}$ (and analogously for matrices $U$ and $V$),
the following partitioning will be employed:
\[
  \mtx{T}^{\si}
  \rightarrow
  \begin{pmatrix}
    \mtx{T}_{11}^{\si} & \mtx{T}_{12}^{\si} \\
    \mtx{T}_{21}^{\si} & \mtx{T}_{22}^{\si}
  \end{pmatrix},
\]
where $\mtx{T}^{\si}_{11}$ is $ib \times ib$.
At the $i$-th iteration, for $i=1,\ldots, p $,
we form matrices $\mtx{T}^{\si}, \mtx{U}^{\si}$ and $\mtx{V}^{\si}$ as follows:

\begin{enumerate}[1.]

\item
Construct an orthogonal matrix $\hat{\mtx{V}}^{\si}_{22}$
such that its leading $b$ columns span approximately the same subspace
as the leading $b$ right singular vectors of $\mtx{T}^{\simo}_{22}$.

\item
Compute the unpivoted QR factorization
of $\mtx{T}^{\simo}_{22} \hat{\mtx{V}}^{\si}_{22}(:,1:b)$ to obtain
\[
  \mtx{T}^{\simo}_{22} \hat{\mtx{V}}^{\si}_{22}(:,1:b) =
  \hat{\mtx{U}}_{22}^{\si} \mtx{R}. \]

\item
Compute the SVD of
$\left(\hat{\mtx{U}}_{22}^{\si}(:,1:b)\right)^*
 \mtx{T}^{\simo}_{22}
 \hat{\mtx{V}}^{\si}_{22}(:,1:b)$,
yielding
\[
  \left( \hat{\mtx{U}}_{22}^{\si}(:,1:b) \right)^*
  \mtx{T}^{\simo}_{22}
  \hat{\mtx{V}}^{\si}_{22}(:,1:b)
  =
  \mtx{U}_{\text{SVD}} \mtx{D} \mtx{V}_{\text{SVD}}^*.
\]

\item
Calculate
$\mtx{V}^{\si}$ with
\[
  \mtx{V}^{\si}
  =
  \begin{pmatrix}
    \mtx{I} & 0 \\
    0 & \hat{\mtx{V}}^{\si}_{22}
  \end{pmatrix}
  \begin{pmatrix}
    \mtx{I} & 0 & 0 \\
    0 & \mtx{V}_{\text{SVD}} & 0 \\
    0 & 0 & \mtx{I}
  \end{pmatrix}.
\]

\item
Calculate $\mtx{U}^{\si}$ with
\[
  \mtx{U}^{\si}
  =
  \begin{pmatrix}
    \mtx{I} & 0 \\
    0 & \hat{\mtx{U}}^{\si}_{22}
  \end{pmatrix}
  \begin{pmatrix}
    \mtx{I} & 0 & 0 \\
    0 & \mtx{U}_{\text{SVD}} & 0 \\
    0 & 0 & \mtx{I}
  \end{pmatrix}.
\]

\item
Calculate $\mtx{T}^{\si}$ with
\[
  \mtx{T}^{\si}
  =
  (\mtx{U}^{\si})^* \mtx{T}^{\simo} \mtx{V}^{\si}.
\]

\end{enumerate}

Once all $\lceil n/b \rceil$ iterations have completed,
we may compute the final factors with
\[
  \begin{aligned}
    \mtx{T} & = \mtx{T}^{(p)}, \\
    \mtx{U} & = \mtx{U}^{(0)} \mtx{U}^{(1)} \cdots \mtx{U}^{(p)}, \\
    \mtx{V} & = \mtx{V}^{(0)} \mtx{V}^{(1)} \cdots \mtx{V}^{(p)}.
  \end{aligned}
\]

This leaves only the question of how the matrix $\hat{\mtx{V}}_{22}^{(i)}$
is formed in step 1 above.
The method, inspired from work in randomized linear algebra
including~\cite{2009_szlam_power,2011_martinsson_randomsurvey,
2006_martinsson_random1_orig}, is the following:

\begin{enumerate}[1.]

\item
Draw a random matrix $\mtx{G}^{\si} \in \mathbb{R}^{(m - ib) \times b}$,
with i.i.d.~entries drawn from the standard normal distribution.

\item
Compute the unpivoted QR factorization of
$ ( (\mtx{T}^{\simo}_{22})^*
     \mtx{T}^{\simo}_{22} )^q
  (\mtx{T}_{22}^{\simo})^*
  \mtx{G}$,
where $q$ is some small nonnegative integer, typically less than three.
The result is
\[
  ( (\mtx{T}^{\simo}_{22})^*
    \mtx{T}^{\simo}_{22} )^q
  (\mtx{T}_{22}^{\simo})^*
  \mtx{G}
  =
  \hat{\mtx{V}}_{22}^{(i)}
  \mtx{R}_{\text{trash}}.
\]

\end{enumerate}

This simple algorithm has been demonstrated to consistently provide
high-quality subspace approximations to the space spanned
by the leading $b$ right singular vectors of $\mtx{T}^{\simo}_{22}$.
It is largely for this reason that \randUTV{} reveals rank comparably
to the SVD.
For details on \randUTV{}, see~\cite{2019_martinsson_randUTV_ACMTOMS}.

\subsection{Executing \texttt{randUTV} out-of-core: an algorithm-by-blocks}
\label{sec:randutv-ooc}

For matrices so large they do not fit in RAM, \randUTV{} requires
significant management of I/O tasks.
If the orthogonal matrices $\mtx{U}$ and $\mtx{V}$ are required, then these
must be stored out-of-core as well.
To implement \randUTV{} efficiently under these constraints,
it is helpful to reorganize the algorithm as an \textit{algorithm-by-blocks}.
Like blocked algorithms, algorithms-by-blocks seek to cast most of
the flops in a factorization in terms
of the highly efficient Level 3 BLAS (Basic Linear Algebra Subprograms).
Unlike blocked algorithms,
the algorithms-by-blocks take maximum advantage of the full main memory
by making all the data blocks being transferred of the same size,
which, besides, makes easier to overlap communication and computation.
All these advantages make an algorithm-by-blocks more efficient.
In the following sections, we present the core technologies behind
the design and implementation of \randUTV{}
as an algorithm-by-blocks.

\subsubsection{Algorithms-by-blocks: an overview.}
\label{sec:abb-overview}

When working with matrices stored in RAM,
\textit{blocked} algorithms can improve performances
by processing multiple columns (or rows) of the input matrix $A$
in each iteration of its main loop.
For instance, some classical factorizations
drive several columns to upper triangular form
in each iteration of the main loop.
This design allows most of the operations to be cast in terms of
the Level 3 BLAS (matrix-matrix operations), and
more specifically in \texttt{xgemm} operations (matrix-matrix products).
As vendor-provided and open-source multithreaded implementations
of the Level 3 BLAS are highly efficient
(with performances close to the peak speed),
blocked algorithms usually offer high performances.
Processing one column (or one row) at a time would require
the employment of the slower matrix-vector operations (Level 2 BLAS)
and much more communication.
Thus, a blocked implementation of \randUTV{} relying largely on
standard calls to parallel BLAS was found to be faster
than the highly optimized MKL CPQR implementation for a shared memory system,
\textit{despite \randUTV{} having a much higher flop count than
the CPQR algorithm}~\cite{2019_martinsson_randUTV_ACMTOMS}.

On the other side,
in blocked algorithms
the amount of data being processed by every iteration varies extremely.
Usually, as the factorization advances,
every call to parallel BLAS must handle an increasing (or a decreasing)
amount of data.
For instance,
to factorize an $n \times n$ matrix with block size $b$,
the first iteration of right-looking algorithms
usually requires the processing of the full matrix,
whereas the last iteration requires just
to process a very small amount of data (in some cases a $b \times b$ block).
When the data is stored in an slow external device,
this extremely high variation in the data being transferred
harms performances
since external devices work optimally only for certain given transfer sizes.
Moreover,
these high variation in the data being transferred and processed
makes an optimal scheduling of I/O operations and computational operations
much more difficult since the cost of the operations varies even much more
(the I/O cost is usually $\mathcal{O}(n^2)$,
whereas the computational cost is usually $\mathcal{O}(n^3)$).
In addition, this high variation also makes a poor use of main memory
either at the beginning or at the end of the factorizations.

We are therefore led to seek a technique other than blocking
to obtain higher performances,
although we will not abandon the strategy of
casting most operations in terms of the Level 3 BLAS.
The key lies in changing the method with which we aggregate
multiple lower-level BLAS flops into a single Level 3 BLAS operation.
Blocked algorithms do this by raising the granularity
of the algorithm's main loop.
The alternative approach, called algorithms-by-blocks,
is to instead raise the granularity of the \textit{data}.
With this method, the algorithm may be designed as if only scalar elements
of the input are dealt with at one time.
Then, the algorithm is transformed into Level 3 BLAS
by conceiving of each scalar as a supmatrix or block of size $b \times b$.
Each scalar operation turns into a matrix-matrix operation, and
operations in the algorithm will, at the finest level of detail,
operate on usually a few (between one and four, but usually two or three)
$b \times b$ blocks.
Each operation on a few blocks is called a task.
This arrangement removes the problems of blocked algorithms
since every task will work with a similar amount of data,
the memory can be employed as a cache to store blocks of memory,
and an overlapping of computation and communication is much easier
and more efficient.
The performance benefits obtained by algorithm-by-blocks
with respect to blocked algorithms
for linear algebra problems on shared-memory architectures
with data stored in RAM are usually significant~\cite{quintana2011oocusing}
when more than a few cores are employed,
because they remove the thread synchronization points
that blocked algorithms insert after every call to the parallel BLAS.
On the other side,
when the data is stored in an external device,
a runtime~\cite{quintana2012oocruntime}
to execute algorithms-by-blocks was described,
but it was only applied to the following basic factorizations:
Cholesky, LU with incremental pivoting, and unpivoted incremental QR.
In our work we have built an algorithm-by-blocks
for computing the much more complex \randUTV{} factorization
on large matrices stored in an external device
by extending the mentioned approach and runtime.
Despite the original \randUTV{} factorization
being much more difficult and complex than those basic factorizations,
our approach to reveal the rank of matrices stored in an external device
has obtained good performances when compared to
high-performance algorithms revealing the rank of matrices stored in RAM,
thus making the revealing of the rank of very large matrices feasible.

An algorithm-by-blocks for computing the \randUTV{}
requires that the QR factorization performed inside
also works on $b \times b$ blocks.
In order to design this internal QR factorization process
such that each unit of work requires only $b \times b$ submatrices,
the algoritm-by-blocks for computing the QR factorization
must employ an algorithm based on updating an existing QR factorization.
We shall refer to this algorithm as \texttt{QR\_AB}.
We consider only the part of \texttt{QR\_AB}
that makes the first column of blocks upper triangular,
since that is all what is required for \texttt{randUTV\_AB}.

Figure~\ref{fig:qr_ab} shows this process for a $9 \times 9$ matrix
with block size 3.
In this figure,
the continuous lines show the $3 \times 3$ blocks of the matrix
involved in the current task,
the `$\bullet$' symbol represents a non-modified element by the current task,
the `$\star$' symbol represents a modified element by the current task, and
the `$\cdot$' symbol represents an element nullified by the current task.
The nullified elements are shown because,
as usual in linear algebra factorizations,
they store information about the Householder transformations
that will be used later to apply these transformations.
The first task, called \textit{Compute\_QR},
computes the QR factorization of the leading dense block $A_{00}$,
thus nullifying all the elements below the main diagonal and
modifying all the elements on or above the main diagonal.
The second task, called \textit{Apply\_left\_Qt\_of\_dense\_QR},
applies the Householder transformations obtained in
the previous task (and stored in $A_{00}$) to block $A_{01}$.
The third task performs the same operation onto $A_{02}$.
The fourth task annihilates block $A_{10}$,
which is called \textit{Compute\_QR\_of\_td\_QR}
(where `td` means triangular-dense
since the upper block $A_{00}$ is triangular and
the lower block $A_{10}$ is dense).
The fifth task,
called \textit{Apply\_left\_Qt\_of\_td\_QR},
applies the transformations of the previous task
to blocks $A_{01}$ and $A_{11}$.
The sixth task performs the same operation onto $A_{02}$ and $A_{12}$.
Analogously,
the seventh, eighth, and ninth tasks update the first and third row of blocks
by performing the same work as the fourth, fifth, and sixth tasks.
By taking advantage of the zeros present in the factorizations for each
iteration, a well-implemented \texttt{QR\_AB}
costs essentially no more flops than the traditional blocked unpivoted QR.
The algorithm is described
in greater detail in~\cite{SuperMatrix:TOMS,SuperMatrix:PPoPP08}.

%
%

\newcommand{\mybl}{\bullet}
\newcommand{\mybn}{\multicolumn{1}{c}{\bullet}} 
\newcommand{\mysl}{\star}
\newcommand{\mysn}{\multicolumn{1}{c}{\star}}
\newcommand{\mypl}{\cdot}
\newcommand{\mypn}{\multicolumn{1}{c}{\cdot}}

\renewcommand{\thesubfigure}{\arabic{subfigure}}

\begin{figure}
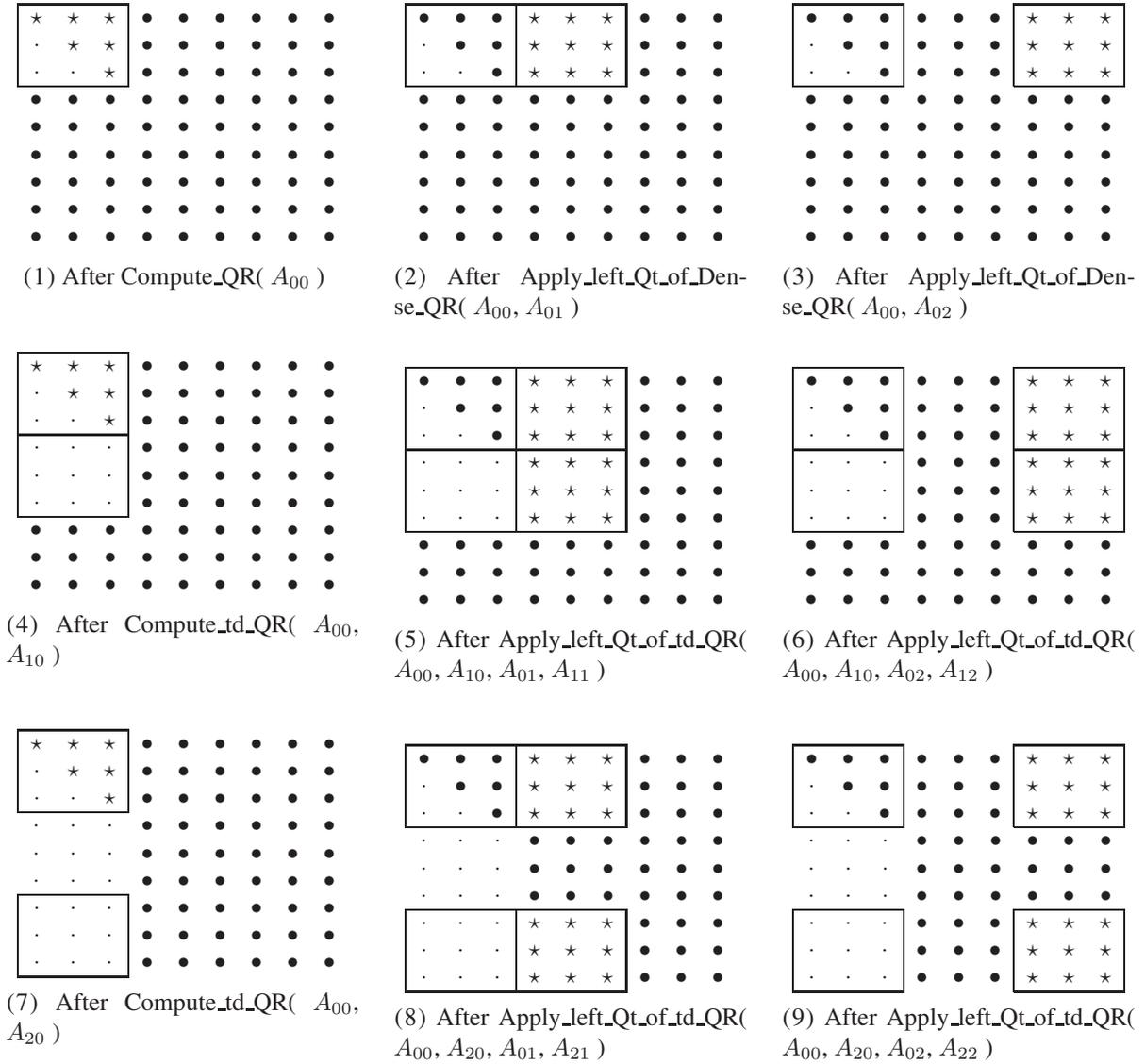


%
%
%
\begin{subfigure}{.30\linewidth}
\footnotesize
\centering
\[
\begin{array}{|ccc|cccccc} 
\cline{1-3}
  \mysl & \mysl & \mysl & \mybl & \mybl & \mybl & \mybl & \mybl & \mybl \\
  \mypl & \mysl & \mysl & \mybl & \mybl & \mybl & \mybl & \mybl & \mybl \\
  \mypl & \mypl & \mysl & \mybl & \mybl & \mybl & \mybl & \mybl & \mybl \\ 
\cline{1-3}
  \mybn & \mybl & \mybn & \mybl & \mybl & \mybl & \mybl & \mybl & \mybl \\
  \mybn & \mybl & \mybn & \mybl & \mybl & \mybl & \mybl & \mybl & \mybl \\
  \mybn & \mybl & \mybn & \mybl & \mybl & \mybl & \mybl & \mybl & \mybl \\
  \mybn & \mybl & \mybn & \mybl & \mybl & \mybl & \mybl & \mybl & \mybl \\
  \mybn & \mybl & \mybn & \mybl & \mybl & \mybl & \mybl & \mybl & \mybl \\
  \mybn & \mybl & \mybn & \mybl & \mybl & \mybl & \mybl & \mybl & \mybl \\
\end{array}
\]
\minusvs
\caption{After Compute\_QR( $A_{00}$ ) \newline}
\vs
\end{subfigure}
\hspace*{0.3cm}
\begin{subfigure}{.30\linewidth}
\footnotesize
\centering
\[
\begin{array}{|ccc|ccc|ccc} 
\cline{1-6}
  \mybl & \mybl & \mybl & \mysl & \mysl & \mysl & \mybl & \mybl & \mybl \\
  \mypl & \mybl & \mybl & \mysl & \mysl & \mysl & \mybl & \mybl & \mybl \\
  \mypl & \mypl & \mybl & \mysl & \mysl & \mysl & \mybl & \mybl & \mybl \\ 
\cline{1-6}
  \mybn & \mybl & \mybn & \mybl & \mybl & \mybn & \mybl & \mybl & \mybl \\
  \mybn & \mybl & \mybn & \mybl & \mybl & \mybn & \mybl & \mybl & \mybl \\
  \mybn & \mybl & \mybn & \mybl & \mybl & \mybn & \mybl & \mybl & \mybl \\
  \mybn & \mybl & \mybn & \mybl & \mybl & \mybn & \mybl & \mybl & \mybl \\
  \mybn & \mybl & \mybn & \mybl & \mybl & \mybn & \mybl & \mybl & \mybl \\
  \mybn & \mybl & \mybn & \mybl & \mybl & \mybn & \mybl & \mybl & \mybl \\
\end{array}
\]
\minusvs
\caption{After Apply\_left\_Qt\_of\_Den\-se\_QR( $A_{00}$, $A_{01}$ )}
\vs
\end{subfigure}
\hspace*{0.3cm}
\begin{subfigure}{.30\linewidth}
\footnotesize
\centering
\[
\begin{array}{|ccc|ccc|ccc|} 
\cline{1-3} \cline{7-9}
  \mybl & \mybl & \mybl & \mybl & \mybl & \mybl & \mysl & \mysl & \mysl \\
  \mypl & \mybl & \mybl & \mybl & \mybl & \mybl & \mysl & \mysl & \mysl \\
  \mypl & \mypl & \mybl & \mybl & \mybl & \mybl & \mysl & \mysl & \mysl \\ 
\cline{1-3} \cline{7-9}
  \mybn & \mybl & \mybn & \mybl & \mybl & \mybn & \mybl & \mybl & \mybn \\
  \mybn & \mybl & \mybn & \mybl & \mybl & \mybn & \mybl & \mybl & \mybn \\
  \mybn & \mybl & \mybn & \mybl & \mybl & \mybn & \mybl & \mybl & \mybn \\
  \mybn & \mybl & \mybn & \mybl & \mybl & \mybn & \mybl & \mybl & \mybn \\
  \mybn & \mybl & \mybn & \mybl & \mybl & \mybn & \mybl & \mybl & \mybn \\
  \mybn & \mybl & \mybn & \mybl & \mybl & \mybn & \mybl & \mybl & \mybn \\
\end{array}
\]
\minusvs
\caption{After Apply\_left\_Qt\_of\_Den\-se\_QR( $A_{00}$, $A_{02}$ )}
\vs
\end{subfigure}
\begin{subfigure}{.30\linewidth}
\footnotesize
\centering
\[
\begin{array}{|ccc|cccccc} 
\cline{1-3}
  \mysl & \mysl & \mysl & \mybl & \mybl & \mybl & \mybl & \mybl & \mybl \\
  \mypl & \mysl & \mysl & \mybl & \mybl & \mybl & \mybl & \mybl & \mybl \\
  \mypl & \mypl & \mysl & \mybl & \mybl & \mybl & \mybl & \mybl & \mybl \\ 
\cline{1-3}
  \mypl & \mypl & \mypl & \mybl & \mybl & \mybn & \mybl & \mybl & \mybn \\
  \mypl & \mypl & \mypl & \mybl & \mybl & \mybn & \mybl & \mybl & \mybn \\
  \mypl & \mypl & \mypl & \mybl & \mybl & \mybn & \mybl & \mybl & \mybn \\
\cline{1-3}
  \mybn & \mybl & \mybn & \mybl & \mybl & \mybn & \mybl & \mybl & \mybn \\
  \mybn & \mybl & \mybn & \mybl & \mybl & \mybn & \mybl & \mybl & \mybn \\
  \mybn & \mybl & \mybn & \mybl & \mybl & \mybn & \mybl & \mybl & \mybn \\
\end{array}
\]
\minusvs
\caption{After Compute\_td\_\-QR( $A_{00}$, $A_{10}$ ) \newline}
\vs
\end{subfigure}
\hspace*{0.3cm}
\begin{subfigure}{.30\linewidth}
\footnotesize
\centering
\[
\begin{array}{|ccc|ccc|ccc} 
\cline{1-6}
  \mybl & \mybl & \mybl & \mysl & \mysl & \mysl & \mybl & \mybl & \mybl \\
  \mypl & \mybl & \mybl & \mysl & \mysl & \mysl & \mybl & \mybl & \mybl \\
  \mypl & \mypl & \mybl & \mysl & \mysl & \mysl & \mybl & \mybl & \mybl \\ 
\cline{1-6}
  \mypl & \mypl & \mypl & \mysl & \mysl & \mysl & \mybl & \mybl & \mybn \\
  \mypl & \mypl & \mypl & \mysl & \mysl & \mysl & \mybl & \mybl & \mybn \\
  \mypl & \mypl & \mypl & \mysl & \mysl & \mysl & \mybl & \mybl & \mybn \\
\cline{1-6}
  \mybn & \mybl & \mybn & \mybl & \mybl & \mybn & \mybl & \mybl & \mybn \\
  \mybn & \mybl & \mybn & \mybl & \mybl & \mybn & \mybl & \mybl & \mybn \\
  \mybn & \mybl & \mybn & \mybl & \mybl & \mybn & \mybl & \mybl & \mybn \\
\end{array}
\]
\minusvs
\caption{After Apply\_left\_Qt\_of\_\-td\_QR( $A_{00}$, $A_{10}$, $A_{01}$, $A_{11}$ )}
\vs
\end{subfigure}
\hspace*{0.3cm}
\begin{subfigure}{.30\linewidth}
\footnotesize
\centering
\[
\begin{array}{|ccc|ccc|ccc|} 
\cline{1-3} \cline{7-9}
  \mybl & \mybl & \mybl & \mybl & \mybl & \mybl & \mysl & \mysl & \mysl \\
  \mypl & \mybl & \mybl & \mybl & \mybl & \mybl & \mysl & \mysl & \mysl \\
  \mypl & \mypl & \mybl & \mybl & \mybl & \mybl & \mysl & \mysl & \mysl \\ 
\cline{1-3} \cline{7-9}
  \mypl & \mypl & \mypl & \mybl & \mybl & \mybl & \mysl & \mysl & \mysl \\
  \mypl & \mypl & \mypl & \mybl & \mybl & \mybl & \mysl & \mysl & \mysl \\
  \mypl & \mypl & \mypl & \mybl & \mybl & \mybl & \mysl & \mysl & \mysl \\
\cline{1-3} \cline{7-9}
  \mybn & \mybl & \mybn & \mybl & \mybl & \mybn & \mybl & \mybl & \mybn \\
  \mybn & \mybl & \mybn & \mybl & \mybl & \mybn & \mybl & \mybl & \mybn \\
  \mybn & \mybl & \mybn & \mybl & \mybl & \mybn & \mybl & \mybl & \mybn \\
\end{array}
\]
\minusvs
\caption{After Apply\_left\_Qt\_of\_\-td\_QR( $A_{00}$, $A_{10}$, $A_{02}$, $A_{12}$ )}
\vs
\end{subfigure}
\begin{subfigure}{.30\linewidth}
\footnotesize
\centering
\[
\begin{array}{|ccc|cccccc} 
\cline{1-3}
  \mysl & \mysl & \mysl & \mybl & \mybl & \mybl & \mybl & \mybl & \mybl \\
  \mypl & \mysl & \mysl & \mybl & \mybl & \mybl & \mybl & \mybl & \mybl \\
  \mypl & \mypl & \mysl & \mybl & \mybl & \mybl & \mybl & \mybl & \mybl \\ 
\cline{1-3}
  \mypn & \mypl & \mypn & \mybl & \mybl & \mybn & \mybl & \mybl & \mybn \\
  \mypn & \mypl & \mypn & \mybl & \mybl & \mybn & \mybl & \mybl & \mybn \\
  \mypn & \mypl & \mypn & \mybl & \mybl & \mybn & \mybl & \mybl & \mybn \\
\cline{1-3}
  \mypl & \mypl & \mypl & \mybl & \mybl & \mybn & \mybl & \mybl & \mybn \\
  \mypl & \mypl & \mypl & \mybl & \mybl & \mybn & \mybl & \mybl & \mybn \\
  \mypl & \mypl & \mypl & \mybl & \mybl & \mybn & \mybl & \mybl & \mybn \\
\cline{1-3}
\end{array}
\]
\minusvs
\caption{After Compute\_td\_\-QR( $A_{00}$, $A_{20}$ ) \newline}
\end{subfigure}
\hspace*{0.3cm}
\begin{subfigure}{.30\linewidth}
\footnotesize
\centering
\[
\begin{array}{|ccc|ccc|ccc} 
\cline{1-6}
  \mybl & \mybl & \mybl & \mysl & \mysl & \mysl & \mybl & \mybl & \mybl \\
  \mypl & \mybl & \mybl & \mysl & \mysl & \mysl & \mybl & \mybl & \mybl \\
  \mypl & \mypl & \mybl & \mysl & \mysl & \mysl & \mybl & \mybl & \mybl \\ 
\cline{1-6}
  \mypn & \mypl & \mypn & \mybl & \mybl & \mybn & \mybl & \mybl & \mybn \\
  \mypn & \mypl & \mypn & \mybl & \mybl & \mybn & \mybl & \mybl & \mybn \\
  \mypn & \mypl & \mypn & \mybl & \mybl & \mybn & \mybl & \mybl & \mybn \\
\cline{1-6}
  \mypl & \mypl & \mypl & \mysl & \mysl & \mysl & \mybl & \mybl & \mybn \\
  \mypl & \mypl & \mypl & \mysl & \mysl & \mysl & \mybl & \mybl & \mybn \\
  \mypl & \mypl & \mypl & \mysl & \mysl & \mysl & \mybl & \mybl & \mybn \\
\cline{1-6}
\end{array}
\]
\minusvs
\caption{After Apply\_left\_Qt\_of\_\-td\_QR( $A_{00}$, $A_{20}$, $A_{01}$, $A_{21}$ )}
\end{subfigure}
\hspace*{0.3cm}
\begin{subfigure}{.30\linewidth}
\footnotesize
\centering
\[
\begin{array}{|ccc|ccc|ccc|} 
\cline{1-3} \cline{7-9}
  \mybl & \mybl & \mybl & \mybl & \mybl & \mybl & \mysl & \mysl & \mysl \\
  \mypl & \mybl & \mybl & \mybl & \mybl & \mybl & \mysl & \mysl & \mysl \\
  \mypl & \mypl & \mybl & \mybl & \mybl & \mybl & \mysl & \mysl & \mysl \\ 
\cline{1-3} \cline{7-9}
  \mypn & \mypl & \mypn & \mybl & \mybl & \mybn & \mybl & \mybl & \mybn \\
  \mypn & \mypl & \mypn & \mybl & \mybl & \mybn & \mybl & \mybl & \mybn \\
  \mypn & \mypl & \mypn & \mybl & \mybl & \mybn & \mybl & \mybl & \mybn \\
\cline{1-3} \cline{7-9}
  \mypl & \mypl & \mypl & \mybl & \mybl & \mybl & \mysl & \mysl & \mysl \\
  \mypl & \mypl & \mypl & \mybl & \mybl & \mybl & \mysl & \mysl & \mysl \\
  \mypl & \mypl & \mypl & \mybl & \mybl & \mybl & \mysl & \mysl & \mysl \\
\cline{1-3} \cline{7-9}
\end{array}
\]
\minusvs
\caption{After Apply\_left\_Qt\_of\_\-td\_QR( $A_{00}$, $A_{20}$, $A_{02}$, $A_{22}$ )}
\end{subfigure}

\caption{An illustration of the first tasks peformed by an algorithm-by-blocks for computing the QR factorization. 
The `$\bullet$' symbol represents a non-modified element by the current task,
`$\star$' represents a modified element by the current task, and
`$\cdot$' represents a nullified element by the current task
(they are shown because they store information 
about the Householder transformations that will be later used to apply them).
The continuous lines surround the blocks involved in the current task.}
\label{fig:qr_ab}
\end{figure}

\subsubsection{Algorithms-by-blocks for \texttt{randUTV}}
\label{sec:abb-randutv}

An algorithm-by-blocks for \randUTV{},
which we will call \texttt{randUTV\_AB},
performs mostly the same operations as the original,
but they are rearranged as many more tasks working on usually square blocks
(except maybe the right-most and bottom-most blocks).
We will discuss in some detail how this plays out in the first step of the
algorithm.
First, choose a block size $b$.
(When working on matrices stored on main memory,
small block sizes such as $b=128$ or $256$ usually work well.
In contrast,
when working on very large matrices stored on an external device,
much larger block sizes such as $b=10{,}240$ must be employed.)
For simplicity, assume $b$ divides both $m$ and $n$ evenly.
Recall that at the beginning of \randUTV{},
$T$ is initialized with $T \defeq A$.
Consider the following partitioning of the matrix $T$:
\[
  T
  \rightarrow
  \left (
    \begin{array}{c | c | c | c}
      T_{11} & T_{12} & \cdots & T_{1N} \\ \hline
      T_{21} & T_{22} & \cdots & T_{2N} \\ \hline
      \vdots & \vdots & \ddots & \vdots \\ \hline
      T_{M1} & T_{M2} & \cdots & T_{MN}
    \end{array}
  \right ) ,
\]
where each supmatrix or block $T_{ij}$ is
$b \times b$, $N = n / b$, and $M = m / b$.
Note that the rest of matrices ($G$, $Y$, $U$, and $V$)
must also be accordingly partitioned.
The submatrices $T_{ij}$ (and those of the rest of matrices)
are treated as the fundamental unit of data in the algorithm,
so that each operation is expressed only in these terms.
For the first step of the algorithm when no orthonormal matrices are built,
for instance:

\begin{enumerate}

\item
\textbf{Constructing $V^{(0)}$:}
The first step, $Y^{(0)} = (T^* T)^q T^* G^{(0)}$,
is broken into several tasks,
each one of which computes the product of two blocks.
In the simplified case where $q=0$,
we have $M \times N$ products of two blocks.
The second step, the QR factorization of $Y^{(0)}$,
uses the \texttt{QR\_AB} algorithm previously described.
Thus, the decomposition of each $Y_i^{(0)}$ is computed separately, and
the resulting upper triangular factor $R^{(0)}$ (stored in $Y^{(0)}_1$)
is updated after each step.
Then, matrix $T$ is updated with the previous transformations
obtained in the factorization of $Y^{(0)}$,
that is, $T \leftarrow T V^{(0)}$.
See,
\textit{e.g.}~\cite{gunter2005parallel,quintana2012oocruntime,SuperMatrix:TOMS}
for details on an approach to a full unpivoted QR factorization.

\item
\textbf{Constructing $U^{(0)}$:}
This step requires an unpivoted QR factorization
of the first column block of $T$.
A similar algorithm to the previous one, \texttt{QR\_AB},
has been employed,
the main difference being that the transformations are applied
from the left int this case.
Then, the rest of the matrix $T$ must be updated
with the previous transformations,
that is, $T \leftarrow (U^{(0)})^* T$.

\item
\textbf{Computing the SVD of $T_{11}$:}
This step is the same one as in \randUTV{} and \texttt{randUTV\_AB}.
In both cases, $T_{11}$ is interacted with as a single unit.
Then, matrix $T$ must be properly updated
with the transformations obtained in this step.

\end{enumerate}

\subsubsection{The FLAME abstraction for implementing algorithm-by-blocks.}
\label{sec:flame}

A nontrivial obstacle to implementing an algorithm-by-blocks is
the issue of programmability.
The FLAME (Formal Linear Algebra Methods Environment)
project~\cite{Gunnels:2001:Flame,Igual:2012:Flame} helps to solve this issue.
FLAME is a framework for designing linear-algebra algorithms.
In this approach the input matrix is viewed as
a collection of submatrices,
basing its loops on re-partitionings of the input data.
The FLAME API~\cite{Bientinesi:2005:Representing}
for the C programming language enables a user
to code high-performance implementations of linear-algebra algorithms
at a high level of abstraction.
Besides, this methodology makes it a natural fit
for use with an algorithm-by-blocks.
Thus, the actual code for the implementation of \texttt{randUTV\_AB} looks
very similar to the written version of the algorithm given in Figure
\ref{fig:FLArandutv}.

\setlength{\arraycolsep}{2pt}
\resetsteps 


\renewcommand{\routinename}{ 
  \left[ U,T,V \right] := \mbox{\sc randUTV\_AB}( A, q, n_b )
}


\renewcommand{\initialize}{
  $
    \begin{array}{@{}l}
      V \defeq \mbox{\sc eye}(n(A), n(A)) \\
      U \defeq \mbox{\sc eye}(m(A), m(A))
    \end{array}
  $
}


\renewcommand{\partitionings}{
  $
  A \rightarrow
  \FlaTwoByTwo{A_{TL}}{A_{TR}}
              {A_{BL}}{A_{BR}}
  $
  ,
  $
  V \rightarrow
  \FlaOneByTwo{V_{L}} {V_{R}}
  $
  ,
  $
  U \rightarrow
  \FlaOneByTwo{U_{L}}
              {U_{R}}
  $
}

\renewcommand{\partitionsizes}{
  $ A_{TL} $ is $ 0 \times 0 $,
  $ V_{L} $ has $ 0 $ columns,
  $ U_{L} $ has $ 0 $ columns
}

\renewcommand{\guard}{
  m( A_{TL} ) < m( A )
}


\renewcommand{\blocksize}{b=\min(n_b,n(A_{BR}))}

\renewcommand{\repartitionings}{
  $  
    \FlaTwoByTwo{A_{TL}}{A_{TR}}
                {A_{BL}}{A_{BR}}
    \rightarrow
    \FlaThreeByThreeBR{A_{00}}{A_{01}}{A_{02}}
                      {A_{10}}{A_{11}}{A_{12}}
                      {A_{20}}{A_{21}}{A_{22}}
  $
  , 
  $
    \FlaOneByTwo{ V_L } { V_R }
    \rightarrow
    \FlaOneByThreeR{V_0} {V_1} {V_2}
  $
  , 
  $  
    \FlaOneByTwo{ U_L } { U_R }
    \rightarrow
    \FlaOneByThreeR{U_0} {U_1} {U_2}
  $
}

\renewcommand{\repartitionsizes}{
  $ A_{11} $ is $ b \times b $,
  $ V_1 $ has $ b $ rows,
  $ U_1 $ has $ b $ rows
}


\renewcommand{\moveboundaries}{
  $  
    \FlaTwoByTwo{A_{TL}}{A_{TR}}
                {A_{BL}}{A_{BR}}
    \leftarrow
    \FlaThreeByThreeTL{A_{00}}{A_{01}}{A_{02}}
                      {A_{10}}{A_{11}}{A_{12}}
                      {A_{20}}{A_{21}}{A_{22}}
  $
  , 
  $
    \FlaOneByTwo{ V_L } { V_R }
    \leftarrow
    \FlaOneByThreeL{V_0} {V_1} {V_2}
  $
  , 
  $  
    \FlaOneByTwo{ U_L } { U_R }
    \leftarrow
    \FlaOneByThreeL{U_0} {U_1} {U_2}
  $
}


\renewcommand{\update}{
$
  \hspace{1em} 
  \begin{array}{lcl}
    \% \text{ Right transform } V \\
    G & \defeq & 
      \mbox{\sc generate\_iid\_stdnorm\_matrix}
                 (m(A) - m(A_{00}), n_b) \\
    Y & \defeq & 
      \left( 
        \FlaTwoByTwoSingleLine{A_{11}}{A_{12}} {A_{21}}{A_{22}}^*
        \FlaTwoByTwoSingleLine{A_{11}}{A_{12}} {A_{21}}{A_{22}} 
      \right)^q
      \FlaTwoByTwoSingleLine{A_{11}}{A_{12}} {A_{21}}{A_{22}}^*
      G \\
    {[}Y,T_V] & \defeq & 
      \mbox{\sc unpivoted\_QR}(Y) \\
    \FlaTwoByTwoSingleLine{A_{11}}{A_{12}} {A_{21}}{A_{22}} & \defeq &
      \FlaTwoByTwoSingleLine{A_{11}}{A_{12}} {A_{21}}{A_{22}} - 
      \FlaTwoByTwoSingleLine{A_{11}}{A_{12}}{A_{21}}{A_{22}} 
      W_V T_V W_V^* \\
    \FlaOneByTwoSingleLine {V_1}{V_2} & \defeq &
      \FlaOneByTwoSingleLine {V_1}{V_2} - 
      \FlaOneByTwoSingleLine {V_1}{V_2} W_V T_V W_V^* \\
    \% \text{ Left transform } U \\
    {[}\FlaTwoByOneSingleLine{A_{11}} {A_{21}},T_U{]} & \defeq & 
       \mbox{\sc unpivoted\_QR}
       \left( \FlaTwoByOneSingleLine {A_{11}} {A_{21}} \right ) \\
    \FlaTwoByOneSingleLine{A_{12}}{A_{22}} & \defeq & 
      \FlaTwoByOneSingleLine{A_{12}}{A_{22}} - 
      W_U^*T_U^*W_U \FlaTwoByOneSingleLine {A_{12}} {A_{22}} \\
    \FlaOneByTwoSingleLine{U_1}{U_2} & \defeq &
      \FlaOneByTwoSingleLine{U_1}{U_2} - 
      \FlaOneByTwoSingleLine {U_1}{U_2} W_U T_U W_U^* \\
    \% \text{ Small SVD} \\
    {[}A_{11}, U_{SVD}, V_{SVD}] & \defeq & \mbox{\sc SVD}(A_{11})\\
    A_{01} & \defeq & A_{01} V_{SVD} \\
    A_{12} & \defeq & U_{SVD}^* A_{12} \\
    V_1 & \defeq & V_1 V_{SVD} \\
    U_1 & \defeq & U_1 U_{SVD}
  \end{array}
$
}

\begin{figure}[tbp]
\FlaAlgorithmWithInit
\caption{The \texttt{randUTV} algorithm adapted for algorithms-by-blocks written with the FLAME methodology/notation.  In this algorithm, $
  W_V$ and $W_U$ are the unit lower trapezoidal matrices stored below the
  diagonal of $ Y $ and $\protect\FlaTwoByOneSingleLine {A_{11}} {A_{21}}$, respectively.}
\label{fig:FLArandutv}
\end{figure}

\subsubsection{Scheduling and dispatching the operations
               for an algorithm-by-blocks.}
\label{sec:scheduler}

The runtime described in~\cite{quintana2012oocruntime}
allows to execute algorithms-by-blocks for single factorizations
that work on data stored
in an external device.
We have employed and extended that runtime to compute the
\randUTV{} factorization.
This algorithm is called the \texttt{randUTV\_AB}.
To understand how this OOC runtime works,
consider the problem of factorizing a matrix of $2 \times 2$ blocks
\[
  A
  \leftarrow
  \FlaTwoByTwoSingleLine {A_{00}} {A_{01}} {A_{10}} {A_{11}},
\]
where each block is of size $b \times b$.
We will consider the case where the power iteration parameter $q=0$
for simplicity.
The execution of the program proceeds in two phases:
the analysis stage and the execution stage.

%
%
In the first stage (analysis),
instead of executing the code sequentially,
the runtime builds a list or queue of tasks
recording the operands associated with each task or operation.
Figure~\ref{fig:analyzer}
shows an example of the list built up by the runtime for \texttt{randUTV\_AB}
for the case
$A \in \mathbb{R}^{n \times n}$, $b=n/2$, and $q=0$.
The S factors obtained in the QR factorization of
the $Y$ blocks
are stored in blocks called $B$,
whereas
the S factors obtained in the factorization of
the blocks of the current column block of $A$
are stored in blocks called $C$.

%
%
In the second stage (execution),
the runtime executes (or dispatches)
all the tasks in the list.
Several method can be employed to execute these tasks.

\begin{itemize}

\item
\textbf{Traditional method.}
This is an straightforward implementation.
For each task,
all the input operands are read from the external device,
then the task is executed with its operands stored in RAM,
and finally all the output operands are written to the external device.
The main advantage is its simplicity,
but it obviously performs many I/O operations.

Figure \ref{fig:dispatcher_trad} illustrates the execution
of the first six tasks
of $\texttt{randUTV\_AB}$ for a matrix $A \in \mathbb{R}^{n \times n}$
with block size $b=n/2$.

\item
\textbf{Method with Cache.}
Since the work to do has been decomposed into many ``small'' tasks,
a fast method to accelerate performances is to store as many blocks
as possible in main memory.
By efficiently using a part of the main memory
as a cache of the blocks stored in the external device,
the number of I/O operations can be reduced.
Since most blocks are of the same size,
the management of the cache is very efficient.
The only blocks with different sizes (smaller) are
those that store the S factors of the unpivoted QR factorizations,
but all blocks will be treated in the same way
to simplify the programming and accelerate the execution.
When the cache is full with blocks and a new block must be loaded,
the Least-Recently Used (LRU) block is selected.
If the block has been modified while staying in RAM,
it must be written to the external device.
Otherwise, it can plainly be discarded and overwritten with the new block.

Figure \ref{fig:dispatcher_cache} illustrates the execution
of the first six tasks
of $\texttt{randUTV\_AB}$ for a matrix $A \in \mathbb{R}^{n \times n}$
with block size $b=n/2$ and a cache with 7 blocks.
To reduce the length of the example
but still show the benefits of this approach,
the matrix is small and the cache of blocks is rather large.
Note that when the cache is full, a block must be replaced.
For instance, to load $A_{11}$ block,
the least-recently used block ($A_{00}$) is selected.
As it has not been modified, it need not be written to disk.

\item
\textbf{Method with Cache + Overlapping of computation and communication.}
Though the use of a cache allows
the reuse of blocks currently stored in main memory,
when a block is not already in main memory,
it must be read from the slow external device.
An additional problem happens if the cache is full and
the block selected to be replaced has been modified,
because it must be written to the external device
before loading the new block.
During all this time, the cores cannot compute
and therefore overall performances drop.
One way to avoid this
is to use one core to perform all the I/O operations (communications)
while the other cores compute.
The disadvantage of this method is
that the computations will be slower than the two previous methods
since they employ one fewer core.
However, the communications can be made transparent (or at least be reduced)
since I/O operations are performed at the same time as the computation.

The I/O thread and the rest of the threads (computational threads)
are completely decoupled.
The I/O thread works by bringing data into the cache of blocks in main memory
in advance, and sometimes it must remove data that is currently in cache
to make room for newer data.
This decoupling makes the programming more difficult,
but it accelerates the execution
since the speed of cores and the speed of the external devices can be very
different between computers (or even between different external devices
in the same computer).

Figure \ref{fig:dispatcher_overlapping} illustrates the execution
with overlapping of computation and communication
of the first six tasks
of $\texttt{randUTV\_AB}$ for a matrix $A \in \mathbb{R}^{n \times n}$
with block size $b=n/2$, and a cache with 7 blocks.
In this case the execution of tasks is not clearly marked
since the execution of different tasks can overlap.
For instance reading or writing of operands of one task
can be executed at the same time as the computation of another task.
For ease of notation,
the figure shows that each I/O operation takes
exactly the same as a computational task.
In practice, the I/O operations and the computational tasks are decoupled.

\end{itemize}

\begin{figure}
\tfvspace
\begin{tabular}{|l|c|c|}
  \hline
  \multicolumn{1}{|c|}{Operation}   &
    \multicolumn{2}{c|}{Operands} \\ \cline{2-3}
    & \multicolumn{1}{c|}{In}   & \multicolumn{1}{c|}{Out} \\ \hline
  Generate\_normal\_random &
    &
    $G_0$ \\ \hline
  Generate\_normal\_random &
    &
    $G_1$ \\ \hline
  Gemm\_tn\_oz: $C = A^* B$ &
    $A_{00}$, $G_{0}$   &
    $Y_0$ \\ \hline
  Gemm\_tn\_oz: $C = A^* B$ &
    $A_{01}$, $G_{0}$   &
    $Y_1$ \\ \hline
  Gemm\_tn\_oo: $C = C + A^* B$ &
    $A_{10}$, $G_{1}$, $Y_0$   &
    $Y_{0}$ \\ \hline
  Gemm\_tn\_oo: $C = C + A^* B$ &
    $A_{11}$, $G_{1}$, $Y_1$ &
    $Y_{1}$ \\ \hline
  Comp\_dense\_QR &
    $Y_0$ &
    $Y_{0}$, $B_0$ \\ \hline
  Comp\_td\_QR &
    $Y_0$, $Y_1$ &
    $Y_0$, $Y_1$, $B_1$ \\ \hline
  Apply\_right\_Q\_of\_dense\_QR &
    $Y_0$, $B_0$, $A_{00}$ &
    $A_{00}$ \\ \hline
  Apply\_right\_Q\_of\_dense\_QR &
    $Y_0$, $B_0$, $A_{10}$ &
    $A_{10}$ \\ \hline
  Apply\_right\_Q\_td\_QR &
    $Y_1$, $B_1$, $A_{00}$, $A_{01}$ &
    $A_{00}$, $A_{01}$ \\ \hline
  Apply\_right\_Q\_td\_QR &
    $Y_1$, $B_1$, $A_{10}$, $A_{11}$ &
    $A_{10}$, $A_{11}$ \\ \hline
  Comp\_dense\_QR &
    $A_{00}$ &
    $A_{00}$, $C_0$ \\ \hline
  Comp\_td\_QR &
    $A_{00}$, $A_{10}$ &
    $A_{00}$, $A_{10}$, $C_1$ \\ \hline
  Apply\_left\_Qt\_of\_dense\_QR &
    $A_{00}$, $C_0$, $A_{01}$ &
    $A_{01}$ \\ \hline
  Apply\_left\_Qt\_of\_td\_QR &
    $A_{10}$, $C_1$, $A_{01}$, $A_{11}$ &
    $A_{01}$, $A_{11}$ \\ \hline
  Keep\_upper\_triang &
    $A_{00}$ &
    $A_{00}$ \\ \hline
  Set\_to\_zero &
    &
    $A_{10}$ \\ \hline
  Svd\_of\_block &
    $A_{00}$ &
    $A_{00}$, $P_0$, $Q_0$ \\ \hline
  Gemm\_abta: $A = B^* A$ &
    $P_0$, $A_{01}$ &
    $A_{01}$ \\ \hline
  Svd\_of\_block &
    $A_{11}$ &
    $A_{11}$, $P_0$, $Q_0$ \\ \hline
  Gemm\_aabt: $A = A B^*$ &
    $A_{01}$, $Q_0$ &
    $A_{01}$ \\ \hline
\end{tabular}
\bfvspace
\caption{A list of the tasks or operations queued up by the runtime during
the analyzer stage in the simplified case that the block size is $b = n/2$.
The ``In'' column specifies pieces of required input data.
The ``Out'' column specifies required pieces of data that will be
altered upon completion of the operation.
\label{fig:analyzer}}
\end{figure}

\begin{figure}
\tfvspace
\begin{tabular}{|l|}
  \multicolumn{1}{|c|}{Task} \\ \hline

  $G_0$ $\leftarrow$ Generate\_normal\_random() \\ \hline
  Write\_block( $G_0$ ) \\ \hline
  \hline

  $G_1$ $\leftarrow$ Generate\_normal\_random() \\ \hline
  Write\_block( $G_1$ ) \\ \hline
  \hline

  $A_{00}$ $\leftarrow$ Read\_block() \\ \hline
  $G_0$ $\leftarrow$ Read\_block() \\ \hline
  $Y_0$ $\leftarrow$ Gemm\_tn\_oz( $A_{00}$, $G_0$ ) \\ \hline
  Write\_block( $Y_0$ ) \\ \hline
  \hline

  $A_{01}$ $\leftarrow$ Read\_block() \\ \hline
  $G_0$ $\leftarrow$ Read\_block() \\ \hline
  $Y_1$ $\leftarrow$ Gemm\_tn\_oz( $A_{01}$, $G_0$ ) \\ \hline
  Write\_block( $Y_1$ ) \\ \hline
  \hline

  $A_{10}$ $\leftarrow$ Read\_block() \\ \hline
  $G_1$ $\leftarrow$ Read\_block() \\ \hline
  $Y_0$ $\leftarrow$ Read\_block() \\ \hline
  $Y_0$ $\leftarrow$ Gemm\_tn\_oo( $Y_0$, $A_{10}$, $G_1$ ) \\ \hline
  Write\_block( $Y_0$ ) \\ \hline
  \hline

  $A_{11}$ $\leftarrow$ Read\_block() \\ \hline
  $G_1$ $\leftarrow$ Read\_block() \\ \hline
  $Y_1$ $\leftarrow$ Read\_block() \\ \hline
  $Y_1$ $\leftarrow$ Gemm\_tn\_oo( $Y_1$, $A_{11}$, $G_1$ ) \\ \hline
  Write\_block( $Y_1$ ) \\ \hline
  \hline

  \multicolumn{1}{|c|}{$\vdots$} \\ \hline
\end{tabular}
\bfvspace
\caption{Execution of the tasks by using the traditional approach.
The execution of the six first tasks
requires 16 I/O operations (6 writes and 10 reads).
Double horizontal lines mark the boundaries between consecutive tasks.
\label{fig:dispatcher_trad}}
\end{figure}

\begin{figure}
\tfvspace
\begin{tabular}{|l|ccccccc|}
  \multicolumn{1}{|c|}{Task} &
  \multicolumn{7}{c|}{Cache after Task} \\ \hline

  $G_0$ $\leftarrow$ Generate\_normal\_random() &
    $G_0$    & --       & --       & --       & --       & --       & --
  \\ \hline
  \hline

  $G_1$ $\leftarrow$ Generate\_normal\_random() &
    $G_0$    & $G_1$    & --       & --       & --       & --       & --
  \\ \hline
  \hline

  $A_{00}$ $\leftarrow$ Read\_block() &
    $G_0$    & $G_1$    & $A_{00}$ & --       & --       & --       & --
  \\ \hline
  $Y_0$ $\leftarrow$ Gemm\_tn\_oz( $A_{00}$, $G_0$ ) &
    $G_0$    & $G_1$    & $A_{00}$ & $Y_0$    & --       & --       & --
  \\ \hline
  \hline

  $A_{01}$ $\leftarrow$ Read\_block() &
    $G_0$    & $G_1$    & $A_{00}$ & $Y_0$    & $A_{01}$ & --       & --
  \\ \hline
  $Y_1$ $\leftarrow$ Gemm\_tn\_oz( $A_{01}$, $G_0$ ) &
    $G_0$    & $G_1$    & $A_{00}$ & $Y_0$    & $A_{01}$ & $Y_1$    & --
  \\ \hline
  \hline

  $A_{10}$ $\leftarrow$ Read\_block() &
    $G_0$    & $G_1$    & $A_{00}$ & $Y_0$    & $A_{01}$ & $Y_1$    & $A_{10}$
  \\ \hline
  $Y_0$ $\leftarrow$ Gemm\_tn\_oo( $ Y_0$, $A_{10}$, $G_1$ ) &
    $G_0$    & $G_1$    & $A_{00}$ & $Y_0$    & $A_{01}$ & $Y_1$    & $A_{10}$
  \\ \hline
  \hline

  $A_{11}$ $\leftarrow$ Read\_block() &
    $G_0$    & $G_1$    & $A_{11}$ & $Y_0$    & $A_{01}$ & $Y_1$    & $A_{10}$
  \\ \hline
  $Y_1$ $\leftarrow$ Gemm\_tn\_oo( $Y_1$, $A_{11}$, $G_1$ ) &
    $G_0$    & $G_1$    & $A_{11}$ & $Y_0$    & $A_{01}$ & $Y_1$    & $A_{10}$
  \\ \hline
  \hline

  \multicolumn{8}{|c|}{$\vdots$} \\ \hline
\end{tabular}
\bfvspace
\caption{Execution of the tasks by using a cache with 7 blocks.
The execution of the six first tasks
requires only 4 I/O operations (all of them reads).
Double horizontal lines mark the boundaries between consecutive tasks.
\label{fig:dispatcher_cache}}
\end{figure}

\begin{figure}
\tfvspace
\begin{tabular}{|l|l|ccccccc|}
  \multicolumn{1}{|c|}{Computational Task} &
  \multicolumn{1}{c|}{I/O Task} &
  \multicolumn{7}{c|}{Cache after Task} \\ \hline

  $G_0$ $\leftarrow$ Generate\_normal\_random() &
  $A_{00}$ $\leftarrow$ Read\_block() &
    $G_0$    & $A_{00}$ & --       & --       & --       & --       & --
  \\ \hline

  $G_1$ $\leftarrow$ Generate\_normal\_random() &
  $A_{01}$ $\leftarrow$ Read\_block() &
    $G_0$    & $A_{00}$ & $G_1$    & $A_{01}$ & --       & --       & --
  \\ \hline

  $Y_0$ $\leftarrow$ Gemm\_tn\_oz( $A_{00}$, $G_0$ ) &
  $A_{10}$ $\leftarrow$ Read\_block() &
    $G_0$    & $A_{00}$ & $G_1$    & $A_{01}$ & $Y_0$    & $A_{10}$ & --
  \\ \hline

  $Y_1$ $\leftarrow$ Gemm\_tn\_oz( $A_{01}$, $G_0$ ) &
  $A_{11}$ $\leftarrow$ Read\_block() &
    $G_0$    & $G_1$    & $A_{10}$ & $Y_0$    & $A_{01}$ & $A_{10}$ & $Y_1$
  \\ \hline

  $Y_0$ $\leftarrow$ Gemm\_tn\_oo( $Y_0$, $A_{10}$, $G_1$ ) &
  &
    $G_0$    & $G_1$    & $A_{10}$ & $Y_0$    & $A_{01}$ & $A_{10}$ & $Y_1$
  \\ \hline

  $Y_1$ $\leftarrow$ Gemm\_tn\_oo( $Y_1$, $A_{11}$, $G_1$ ) &
  &
    $G_0$    & $G_1$    & $A_{10}$ & $Y_0$    & $A_{01}$ & $A_{10}$ & $Y_1$
  \\ \hline

  \multicolumn{9}{|c|}{$\vdots$} \\ \hline
\end{tabular}
\bfvspace
\caption{Execution of the tasks by using a cache with 7 blocks and
overlapping of computation and communication.
The execution of the six first tasks
requires only 4 I/O operations (all of them reads),
and all of them are peformed at the same time as the computation.
No double horizontal lines are employed to mark
the boundaries between consecutive tasks
since I/O operations and computations of different tasks can be
executed simultaneously.
\label{fig:dispatcher_overlapping}}
\end{figure}

\section{Numerical results}
\label{sec:num}

In this section,
we present the experiments demonstrating the scalability and
computational costs of implementations of
\texttt{HQRRP} and \texttt{randUTV\_AB} for matrices stored out-of-core.
In subsection~\ref{sec:ooc-cpqr-times},
we compare several implementations of \texttt{HQRRP}
with different strategies for handling the I/O.
In subsection~\ref{sec:ooc-randutv-times},
we examine computational cost of an implementation of \randUTV{}.

\subsection{Partial CPQR factorization with \texttt{HQRRP}}
\label{sec:ooc-cpqr-times}

The machine used for these experiments had four DIMM memory chips with
16 GB of DDR4 memory each.
The CPU was an Intel \circledR Core \texttrademark i7-6700K (4.00 GHz) with
four cores.
Experiments were run on two different hard drives.
One was a Toshiba P300 HDD with 3 TB of memory and 7200 RPM;
the other was a Samsung V-NAND SSD 950 Pro with 512 GB of memory.
The code was compiled with \texttt{gcc} (version 5.4.0)
and linked with the Intel \circledR MKL library (Version 2017.0.3).

The computational cost of three different implementations of \texttt{HQRRP}
for matrices stored out-of-core were assessed.

\begin{itemize}

\item
\textbf{Algorithm 1 -- In place:}
This implementation of the \texttt{HQRRP}
did not carry out physical permutations on the columns of $\mtx{A}$
but instead applied the permutation information during each I/O task.
In other words, the routine returns $\mtx{R} \mtx{P}^*$ rather than $\mtx{R}$.
As a result, no more data than necessary is transferred to and
from the hard drive, but many reads occur from noncontiguous locations in
the drive.

\item
\textbf{Algorithm 2 -- Physical pivoting:}
Permutations \textit{are} physically performed during the
computation, so that upon completion $\mtx{R}$ is stored in the hard drive.
This implementation reads and writes more data than the ``in place'' version,
but the data transfer mostly occurs in contiguous portions of memory.

\item
\textbf{Algorithm 3 -- Left-looking:}
An implementation of a left-looking version of \texttt{HQRRP},
outlined in Section \ref{sec:hqrrp-ooc}.
This version requires only $\mathcal{O}(n^2)$ write operations,
but the total number of operations (including reading, writing, and flops)
is asymptotically higher than the ``in place'' and ``physical pivoting''
implementations.

\end{itemize}

\begin{figure}
\includegraphics[width=.55\textwidth]{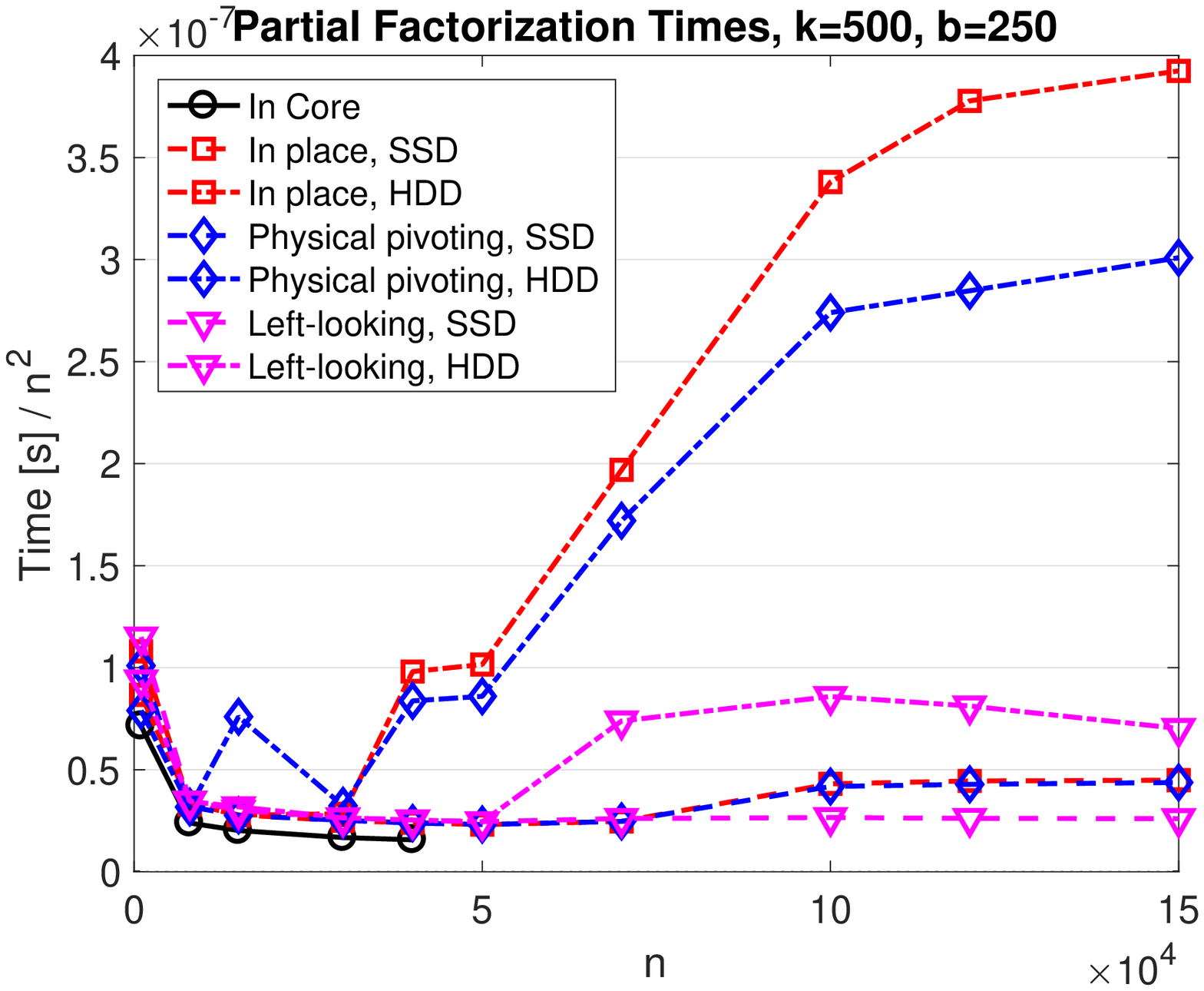}
\includegraphics[width=.55\textwidth]{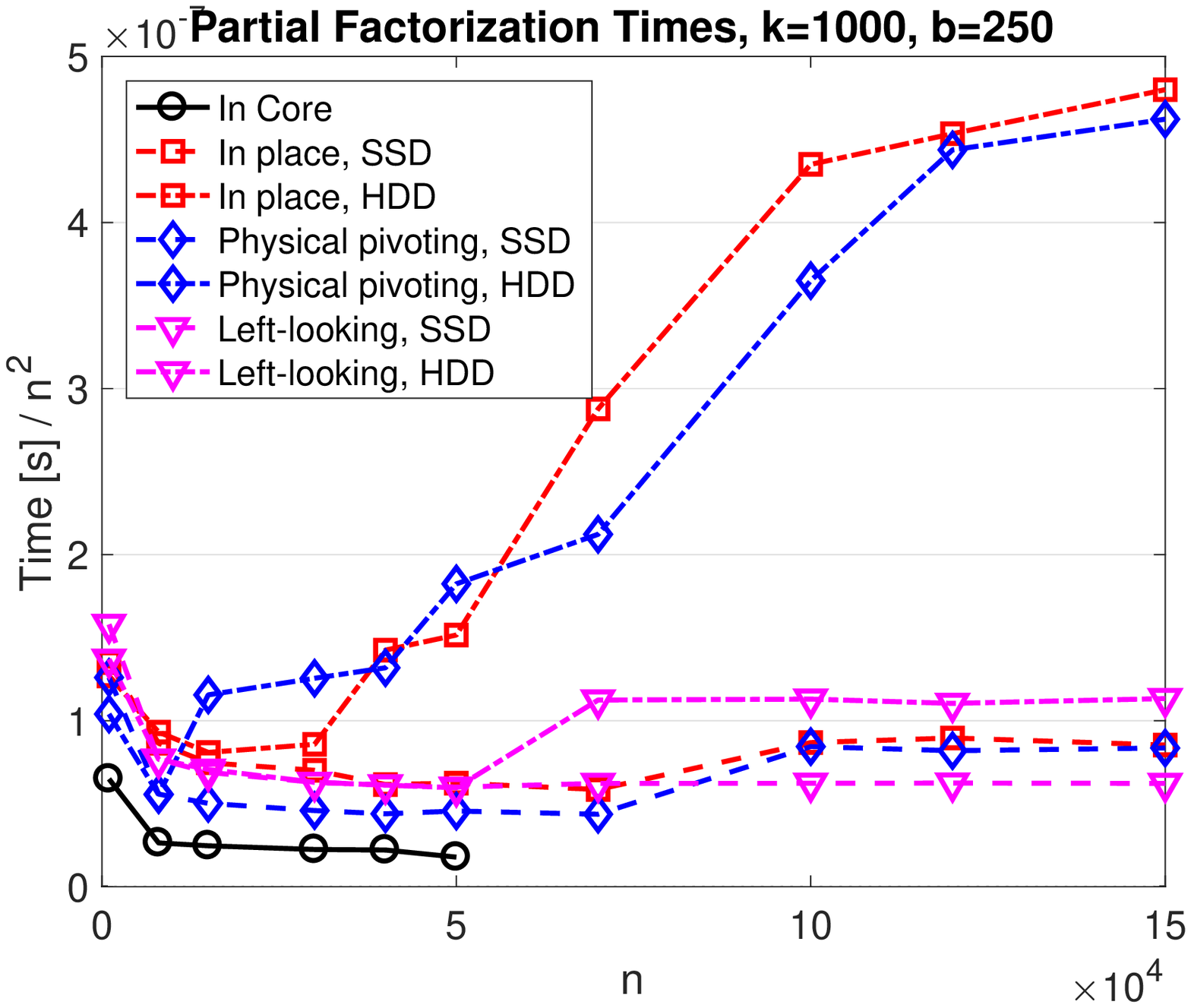}
\caption{A comparison of the computational cost for three different algorithms
for computing a partial CPQR when the matrix may be too large to fit in RAM.
In the top figure, $1000$ columns of the input matrix were processed.
In the bottom figure, $500$ columns were processed.
The block size $b$ in each case was $250$.
\label{fig:ooc-cpqr-partial-times}}
\end{figure}

Figure~\ref{fig:ooc-cpqr-partial-times}
shows the times scaled by the square of the matrix dimension
of several factorizations.
Two key observations can be made of the results depicted in this figure.
The first observation is that
the number of writes required by the implementation has a dramatic
effect on its performance, especially
when the matrix is stored on a hard disk drive.
Algorithms 2 and 3, while performing reasonably well on the SSD,
did not even scale correctly on the HDD, or at the very least did not reach
asymptotic behavior as early as the SSD experiments.
Algorithm 3, which requires $\mathcal{O}(n^2)$ write operations
rather than $\mathcal{O} (n^3)$,
outperforms the right-looking alternatives
on both hard drives for large matrices.
The second observation is that
the best-performing algorithm takes roughly $3$ times longer than
the predicted performance for an in-core factorization of
a matrix of the same size on the SSD.
On the HDD, it takes about $5.4$ times longer.

\subsection{Full factorization with \texttt{randUTV}}
\label{sec:ooc-randutv-times}

In this subsection,
we assess the performances of our new implementations
for computing the \randUTV{} factorization
when the data are stored in a disk drive,
and compare it to the performances of highly optimized methods
for computing the SVD, the column pivoted QR (CPQR) factorization,
and the \randUTV{} factorization when the data are stored in RAM.

To fairly compare the different implementations included in this study,
the usual flop rate or the usual flop count cannot be employed
since the computation of the SVD, the CPQR, and the \randUTV{} factorizations
require a very different number of flops
(the dominant $n^3$-term in the asymptotic flop count is very different).
Absolute computational times are not shown either
as they vary greatly because of the huge range of matrix dimensions
employed in the experiments.
Therefore, scaled computational times
(absolute computational times divided by $n^3$) are employed instead.
The lower the scaled computational times,
the better the performances are.
Since all the implementations being assessed
have asymptotic complexity $O(n^{3})$
when applied to an $n\times n$ matrix,
these graphs better reveal the computational speed.
The scaled times are multiplied by a constant
to make the figures in the vertical axis more readable.
The value of this constant is shown in the vertical axis,
and it is usually $10^{10}$.

For the implementations of \randUTV{}, both in-core and out-of-core,
results are shown for
0, 1, and 2 iterations ($q=0$, $q=1$, and $q=2$, respectively)
in the power iteration process.
Recall that the higher $q$,
the higher the approximation to the singular values obtained in
the main diagonal of matrix $T$.


The following out-of-core implementations were assessed
in the experiments of this subsection:

\begin{itemize}

\item
{\sc OOC-randUTV-T}:
This is the traditional implementation
for computing the \randUTV{} factorization
of a matrix stored in the disk
by using an algorithm-by-blocks.

\item
{\sc OOC-randUTV-V}:
This is the implementation
for computing the \randUTV{} factorization
of a matrix stored in the disk
by using an algorithm-by-blocks
with a cache of blocks and overlapping of computation and I/O.
Unlike the previous implementation,
this one uses all the cores but one for computation and one core for I/O.
Not all of the RAM in the computer is employed in the cache for matrix blocks,
and a large amount of memory is left available for the operating system,
since some initial experiments supported this approach.

\item
{\sc OOC-QR-T}:
This is the traditional right-looking implementation
for computing the QR factorization
of a matrix stored in the disk
by using an algorithm-by-blocks.

\item
{\sc OOC-QR-V}:
This is the right-looking implementation
for computing the QR factorization
of a matrix stored in the disk
by using an algorithm-by-blocks
with a cache of blocks and overlapping of computation and I/O.
One core is employed for I/O and the rest, for computations.
Not all of the RAM in the computer is employed in the cache of matrix blocks.

\end{itemize}

Although the two methods for computing the QR factorization
included in the experiments do not reveal the rank,
they were included as a performance reference for the others.

In order to be included in this study,
we asked some authors to send us their out-of-core codes
to reveal the rank (e.g.~SVD).
Unfortunately, no codes were made available to us.


Our aim is to factorize very large matrices that do not usually fit in RAM
unless a very expensive main memory is available.
However,
as a performance reference for our out-of-core implementations,
we have included the following in-core implementations:

\begin{itemize}

\item
{\sc MKL SVD}:
The routine called \texttt{dgesvd} from MKL's LAPACK
was used to compute the Singular Value Decomposition
of matrices stored in RAM.

\item
{\sc MKL CPQR}:
The routine called \texttt{dgeqp3} from MKL's LAPACK
was used to compute the column-pivoting QR factorization
of matrics stored in RAM.

\item
{\sc randUTV PBLAS}
(\randUTV{} with parallel BLAS):
This is the traditional implementation
for computing the \randUTV{} factorization
of matrics stored in RAM
that relies on the parallel BLAS
to take advantage of all the cores in the system.
The parallel BLAS library from MKL was employed with these codes
for the purpose of a fair comparison.

\item
{\sc randUTV AB}
(\randUTV{} with Algorithm-by-Blocks):
This is the new implementation for computing the \randUTV{} factorization
by scheduling all the tasks to be computed in parallel,
and then executing them with serial BLAS.
The serial BLAS library from MKL was employed with these new codes
for the purpose of a fair comparison.

\item
{\sc MKL QR}:
The routine called \texttt{dgeqrf} from MKL's LAPACK
was used to compute the QR factorization
of matrices stored in RAM.
Although this routine does not reveal the rank,
it was included in some experiments as a performance reference for the others.

\end{itemize}

For most of the experiments, two plots are shown.
The left plot shows the performances
when no orthonormal matrices are computed.
In this case,
just the upper triangular factor $R$ is computed for the CPQR and QR,
just the upper triangular factor $T$ is computed for the \randUTV{},
and just the singular values are computed for the SVD.
In contrast, the right plot shows the performances
when all orthonormal matrices are explicitly formed
in addition to the previously mentioned factors.
In this case,
matrix $Q$ is computed for the CPQR and QR, and
matrices $U$ and $V$ are computed for the \randUTV{} and for the SVD.
The right plot slightly favors the CPQR and QR
since only one orthonormal matrix is formed.

\subsubsection{Experimental Setup}
\label{sec:expsetup}

The experiments reported in this section were performed on three computers.
Next they are briefly described.

\begin{itemize}

\item
\texttt{ua}:
This HP computer contained
two Intel Xeon\circledR\ CPU X5560 processors at 2.8 GHz,
with 12 cores and 48 GiB of RAM in total.
Its OS was GNU/Linux (Version 3.10.0-514.21.1.el7.x86\_64).
Intel's \texttt{icc} compiler (Version 12.0.0 20101006) was employed.
LAPACK and BLAS routines were taken from
the Intel(R) Math Kernel Library (MKL)
Version 10.3.0 Product Build 20100927 for Intel(R) 64 architecture,
since this library usually delivers much higher performances
than LAPACK and BLAS from the Netlib repository.
The Hard Disk Drive (HDD) employed in the experiments
to store all the data in the out-of-core implementations
was an HP MM0500EANCR (Firmware Revision HPG3).
Though its capacity was 500 GB, only about 400 GB were available for users,
which was about 8.3 times as large as the main memory.
According to the Linux operating system \texttt{hdparm} tool,
the read speed of this drive was 91.13 MB/s.

\item
\texttt{ut}:
This Dell computer contained
two Intel Xeon\circledR\ Gold 6254 processors at 3.10GHz,
with 36 cores and 754 GiB of RAM in total.
Its OS was GNU/Linux (Version 5.0.0-37-generic).
Intel's \texttt{icc} compiler (Version 19.0.5.281 20190815) was employed.
LAPACK and BLAS routines were taken from
the Intel(R) Math Kernel Library (MKL)
Version 2019.0.5 Product Build 20190808 for Intel(R) 64 architecture
for the same reason as before.
The disk drive (SDD) employed in the experiments
to store all the data in the out-of-core implementations
was a Toshiba KXG50PNV2T04 NVMe (Firmware Revision AFDA4105)
with a capacity of 2 TiB.
According to the Linux operating system \texttt{hdparm} tool,
the read speed of this disk was 9,744.21 MB/s on cached reads and
the read speed of this disk was 2,410.66 MB/s on buffered disk reads.

\item
\texttt{ucm}:
This Supermicro computer contained
two Intel Xeon\circledR\ processors E5-2695 v3 at 2.30GHz,
with 28 cores and 125 GiB of RAM in total.
In this computer the so-called \textit{Turbo Boost} mode of the CPU
was turned off in our experiments.
Its OS was GNU/Linux (Version 2.6.32-504.el6.x86\_64).
Intel's \texttt{icc} compiler (Version 18.0.1 20171018) was employed.
LAPACK and BLAS routines were taken from
the Intel(R) Math Kernel Library (MKL)
Version 2018.0.1 Product Build 20171007 for Intel(R) 64 architecture
for the same reason as before.
The disk drive (SDD) employed in the experiments
to store all the data in the out-of-core implementations
was a Samsung SSD 850 EVO (Firmware Revision EMT02B6Q)
with a capacity of 1 TB.
According to the Linux operating system \texttt{hdparm} tool,
the read speed of this disk was 10,144.81 MB/s on cached reads and
the read speed of this disk was 441.48 MB/s on buffered disk reads.

\end{itemize}

In our out-of-core implementations the block cache employed
16 GiB (of the 48 GiB),
256 GiB (of the 754 GiB), and
32 GiB (of the 128 GiB)
in
\texttt{ua},
\texttt{ut}, and
\texttt{ucm},
respectively.
The rest was left for the operating system kernel and buffers,
and the application's code.

In contrast,
unless explicitfly stated otherwise,
all the experiments employed all the cores in the computer.

Notice that the \texttt{ua} computer has a very slow spinning disk
as well as a low computational power.
Notice also that both \texttt{ut} and \texttt{ucm} have SSD disks,
but their performances are different.
The \texttt{ucm} computer has an SSD with a SATA interface,
which is limited to 600 MiB/s,
whereas
the \texttt{ut} computer has an SSD with an M.2 interface,
which does not have that limitation.

In all the experiments double-precision real matrices were processed.
All the matrices used in these experiments were randomly generated
since generation is fast.

\subsubsection{Effect of block sizes}


\begin{figure}[ht!]
\tfvspace
\begin{center}
\begin{tabular}{cc}
\includegraphics[width=0.45\textwidth]{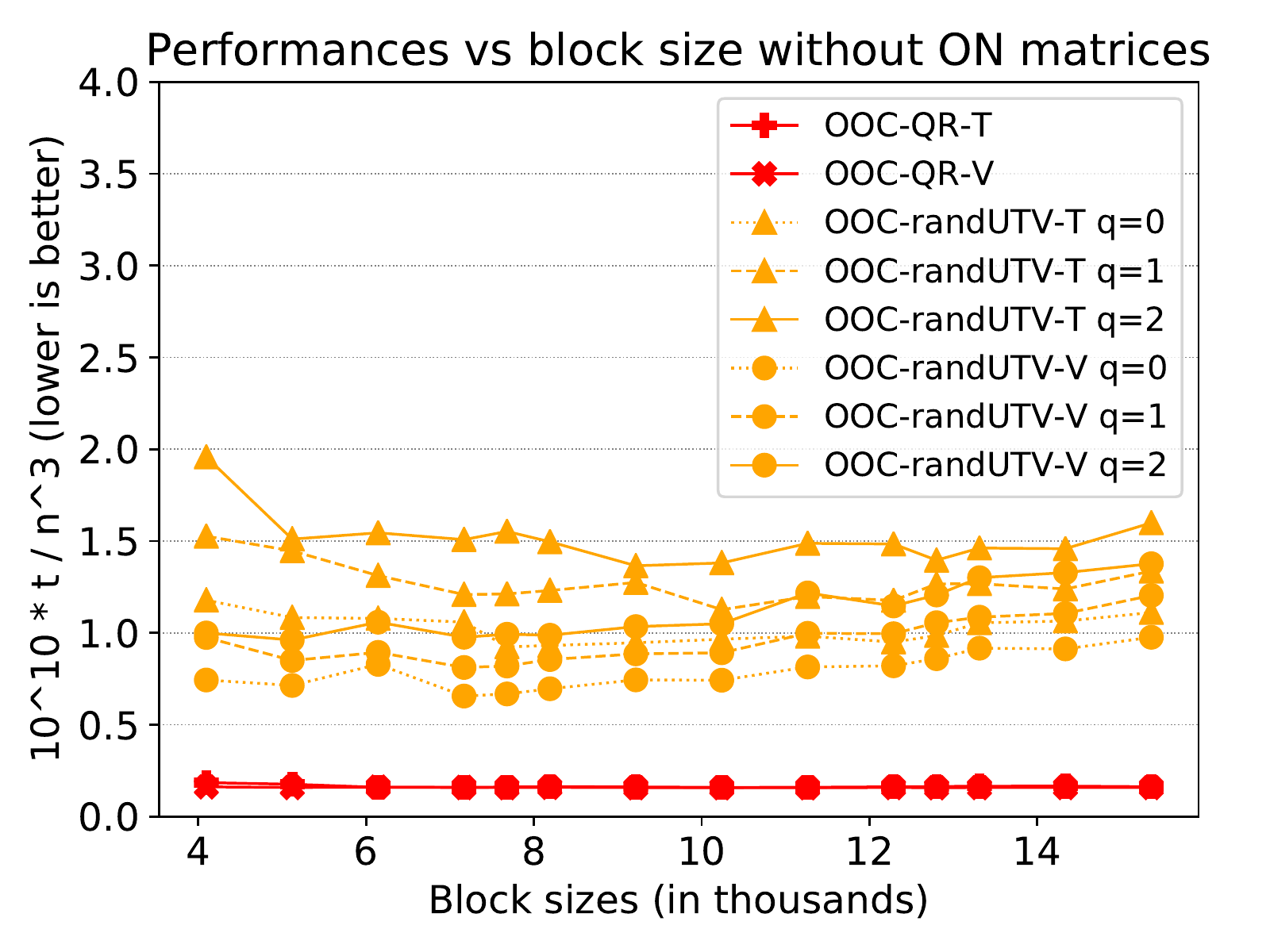} &
\includegraphics[width=0.45\textwidth]{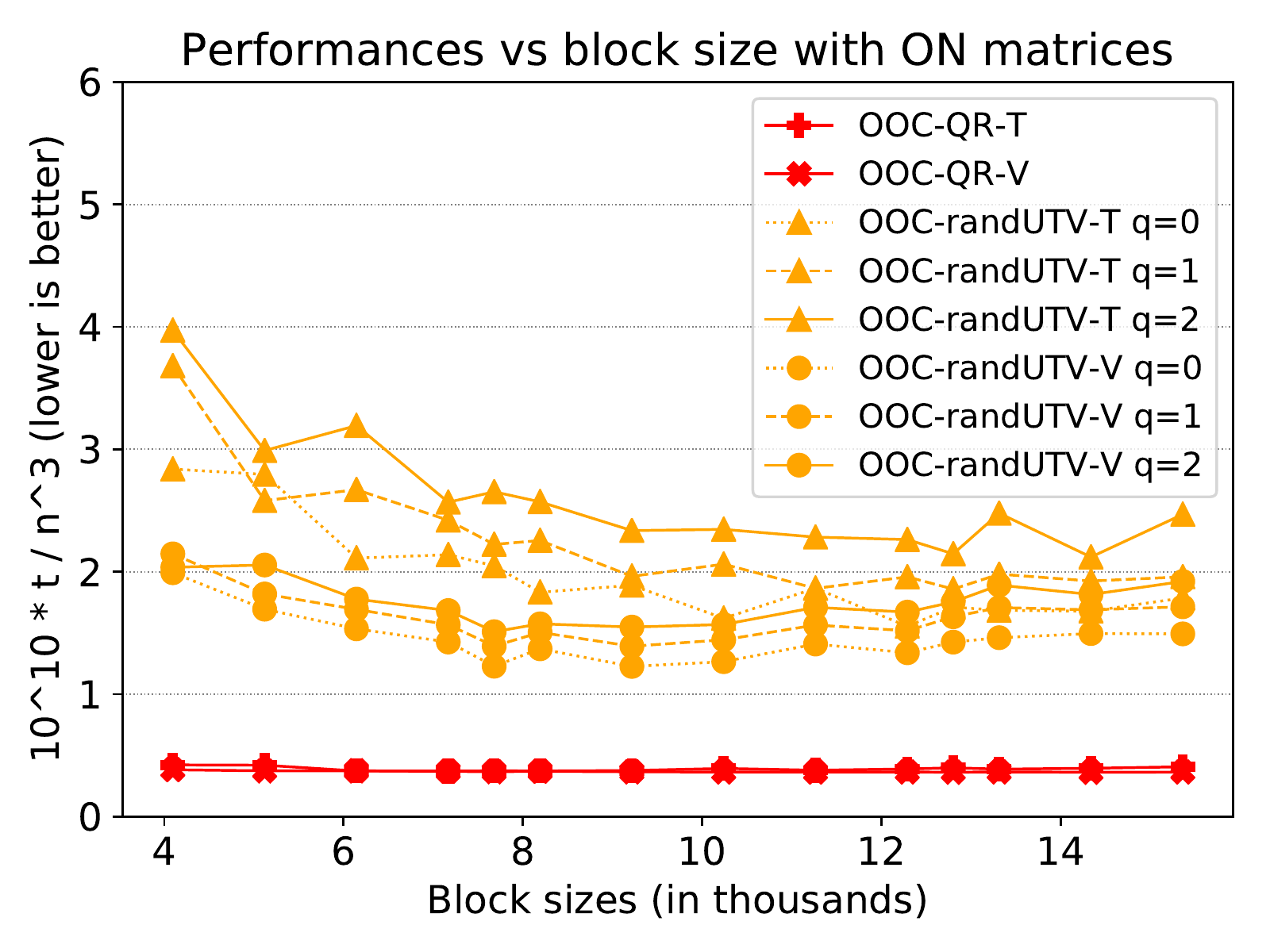} \\
\end{tabular}
\end{center}
\bfvspace
\caption{Performances of QR and \randUTV{} implementations versus block sizes
on matrices of dimension $81{,}920 \times 81{,}920$ in \texttt{ua}.}
\label{fig:ooc_block_size_ua}
\end{figure}

Figure~\ref{fig:ooc_block_size_ua}
shows the scaled computational times
obtained by our implementations for computing 
both the QR factorization and the \randUTV{} factorization
versus several block sizes
when matrices of dimension $81{,}920 \times 81{,}920$ are processed
in \texttt{ua}.
The aim of these two plots is to determine the optimal block sizes.
As can be seen,
performances of the QR factorization
do not strongly depend on the block size.
In contrast,
for the \randUTV{} factorization
performances do depend on the block size,
and block sizes between 7,680 and 11,264 usually offer optimal results.

From now on,
in \texttt{ua}
the block size 10,240 will be employed
for both the out-of-core QR and the out-of-core \randUTV{} factorizations
since it returns near-optimal results in most cases.
As \texttt{ut} and \texttt{ucm} have larger central memories than \texttt{ua},
larger matrices will be assessed in those two computers.
Since in linear algebra
the larger the matrix sizes being tested,
the larger the optimal block sizes usually are,
in \texttt{ut} and \texttt{ucm} the block size 20,480 will be employed.

\subsubsection{Comparison of Out-Of-Core variants}

The plots in all the next figures include a vertical dashed gray line
showing the largest theoretical matrix size that can be stored in the RAM
of the computer when the \randUTV{} factorization is computed.
For instance, in \texttt{ua}
this size is about 80,000 when no orthonormal matrices are built,
and therefore matrices $U$ and $V$ do not need to be stored.
In contrast,
this size is about 46,000 when orthonormal matrices are built,
and therefore matrices $U$, and $V$ must be stored.
%
In practice, the actual threshold must be sligthly smaller
than those in the pictures
since main memory must be also employed
for the operating system kernel,
operating system's disk cache and buffers,
application's code,
application's other data, etc.


\begin{figure}[ht!]
\tfvspace
\begin{center}
\begin{tabular}{cc}
\includegraphics[width=0.45\textwidth]{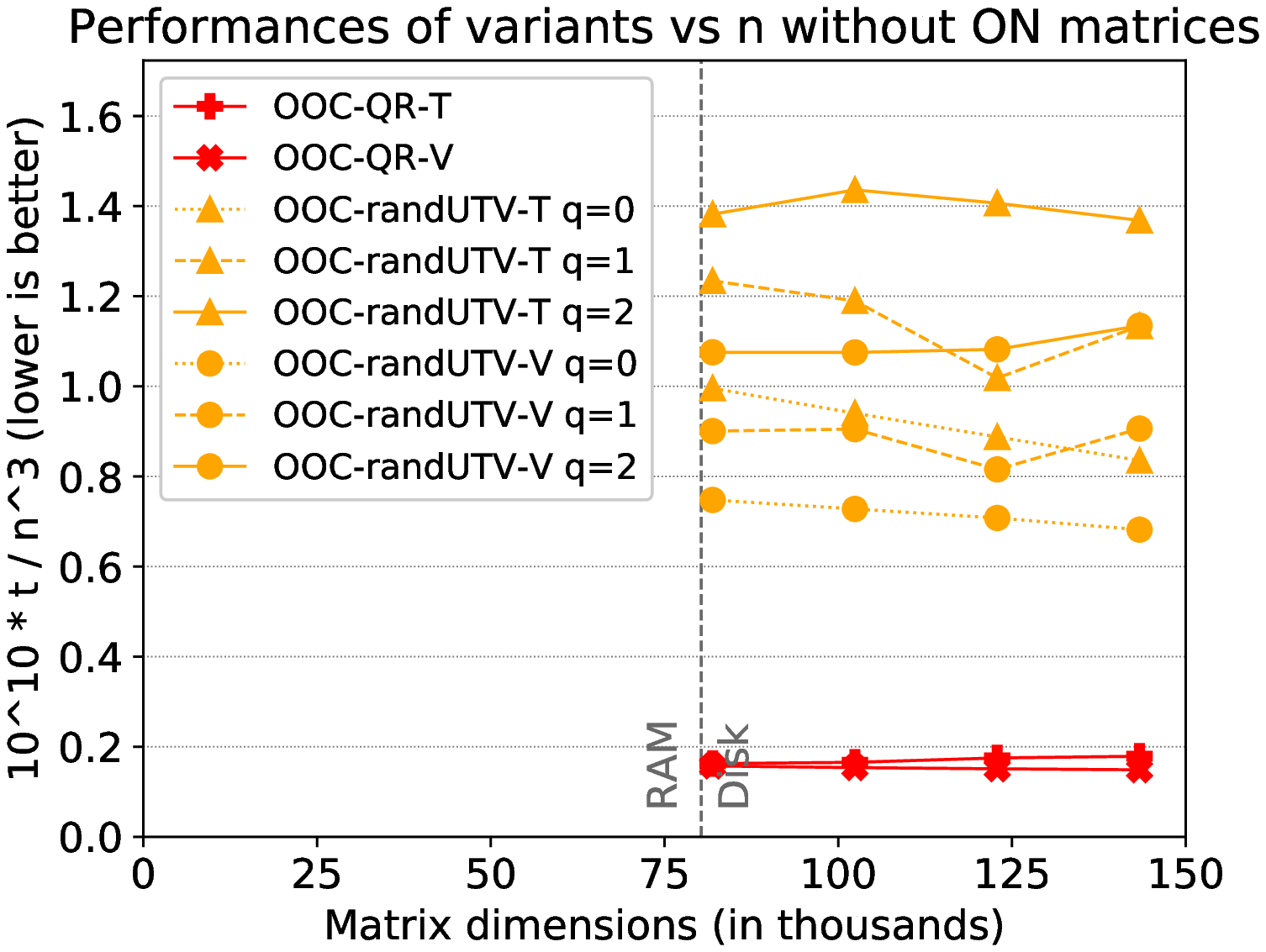} &
\includegraphics[width=0.45\textwidth]{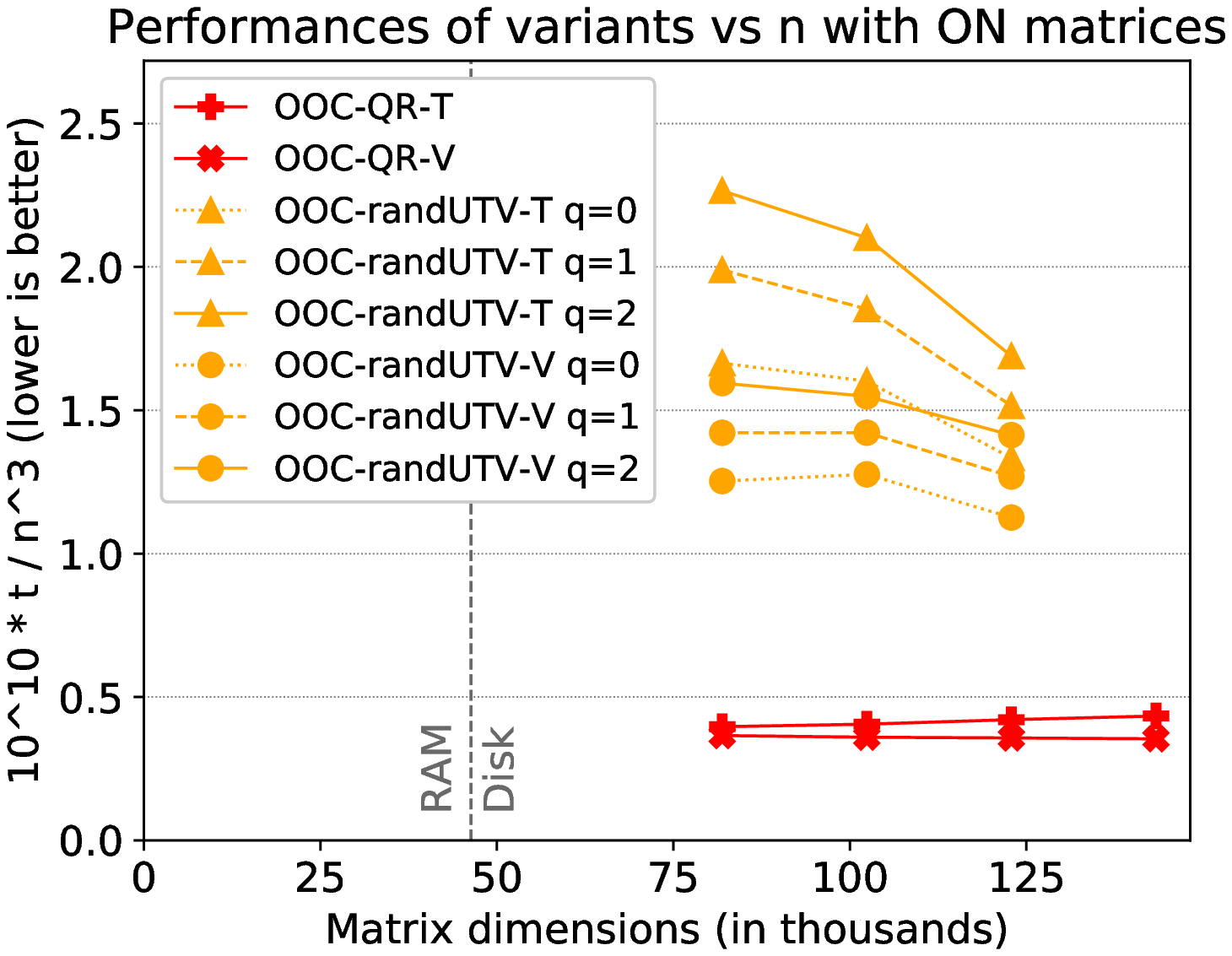} \\
\end{tabular}
\end{center}
\bfvspace
\caption{Performances versus matrix dimensions
for the different implementations of the QR and \randUTV{} factorizations
in \texttt{ua}.}
\label{fig:ooc_variants_ua}
\end{figure}

Figure~\ref{fig:ooc_variants_ua}
shows the performances
for the different implementations of the QR and \randUTV{} factorizations
on several large matrix sizes.
As can be seen,
the overlapping of computation and I/O
for the QR factorization increases performances slightly:
The speedup is about between 3 \% and 20 \%
when no orthonormal matrices are built,
and it is about between 8 \% and 22 \%
when orthonormal matrices are built.
In contrast,
the overlapping of computation and I/O
for the \randUTV{} clearly improve performances:
The speedup is about between 25 \% and 37 \%
when no orthonormal matrices are built,
and it is about between 18 \% and 42 \%
when orthonormal matrices are built.


\begin{table}[ht!]
\vspace*{0.4cm}
\begin{center}
\begin{tabular}{lrr}
  \toprule
  \multicolumn{1}{c}{} &
  \multicolumn{1}{c}{\texttt{ut}} &
  \multicolumn{1}{c}{\texttt{ucm}} \\
  \multicolumn{1}{c}{} &
  \multicolumn{1}{c}{$n=348{,}160$} &
  \multicolumn{1}{c}{$n=143{,}360$} \\ \midrule
    Computational time                   &  229,627.6 &  34,165.5 \\
    I/O time                             &   41,312.9 &  38,253.5 \\
    Computational time + I/O time        &  270,940.5 &  72,419.0 \\ \midrule
    Real time                            &  229,910.0 &  43,326.0 \\ \midrule
    Ratio Real time / Computational time &  1.001     &  1.268  \\
    Ratio Computational time / I/O time  &  5.558     &  0.893  \\
  \bottomrule
\end{tabular}
\end{center}
\caption{Decomposed times (in seconds) of the \randUTV{} factorization
of large matrices of dimension $n \times n$
on the \texttt{ut} and \texttt{ucm} platforms.}
\label{tab:overlapping_ut}
\end{table}

To check the benefits of
the overlapping of computation and I/O in the other platforms,
Table~\ref{tab:overlapping_ut} reports the total and decomposed times
of the computation of
the \randUTV{} factorization of large matrices that do not fit in main memory
in both \texttt{ut}
and \texttt{ucm}
when no orthornormal matrices are built and
$q=0$ (the power iteration process).
To build this table,
each operation in the factorization (both computational and I/O)
was recorded.
As can be seen, in \texttt{ut}
the performances are very encouraging
because the I/O time is much smaller than the computational time
(about 18 \%),
thus making the application computation-bound.
However, in \texttt{ucm} the results are different
since the disk of \texttt{ut} is 5.4 times as fast as the disk of \texttt{ucm}.
The slower disk of \texttt{ucm}
makes that the overall computational time and the overall I/O time
are similar, the latter being slightly larger.
On the other side, in \texttt{ut} the overlapping of computation and I/O is
perfect since the real time is only 0.12 \% larger than the
computational time.
In \texttt{ucm} the overlapping of computation and I/O is
almost perfect since the real time is only 27 \% larger than the
computational time, and 13 \% larger than the I/O time
(in this case larger than the computational time).


\begin{figure}[ht!]
\tfvspace
\begin{center}
\begin{tabular}{cc}
\includegraphics[width=0.45\textwidth]{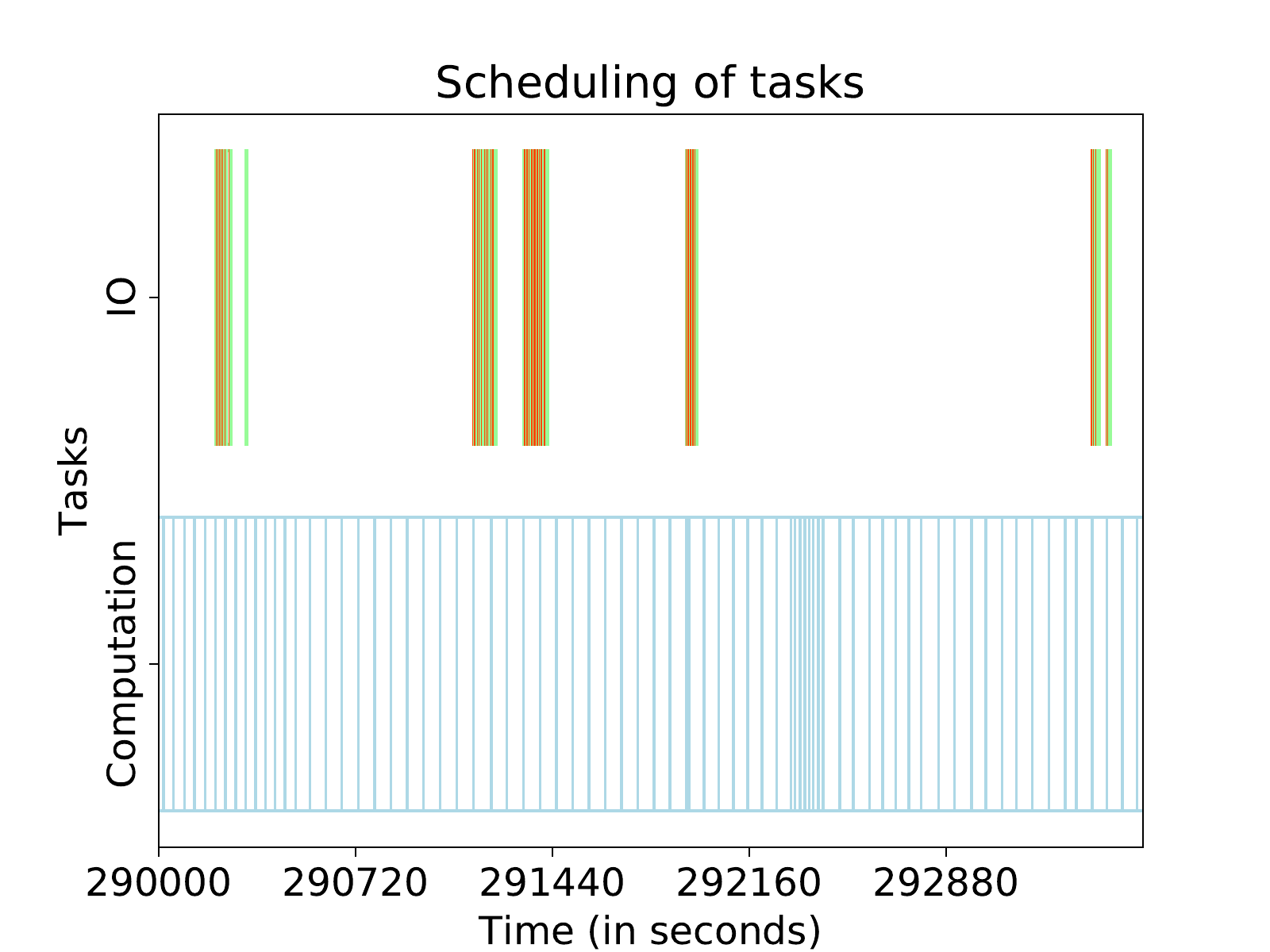} &
\includegraphics[width=0.45\textwidth]{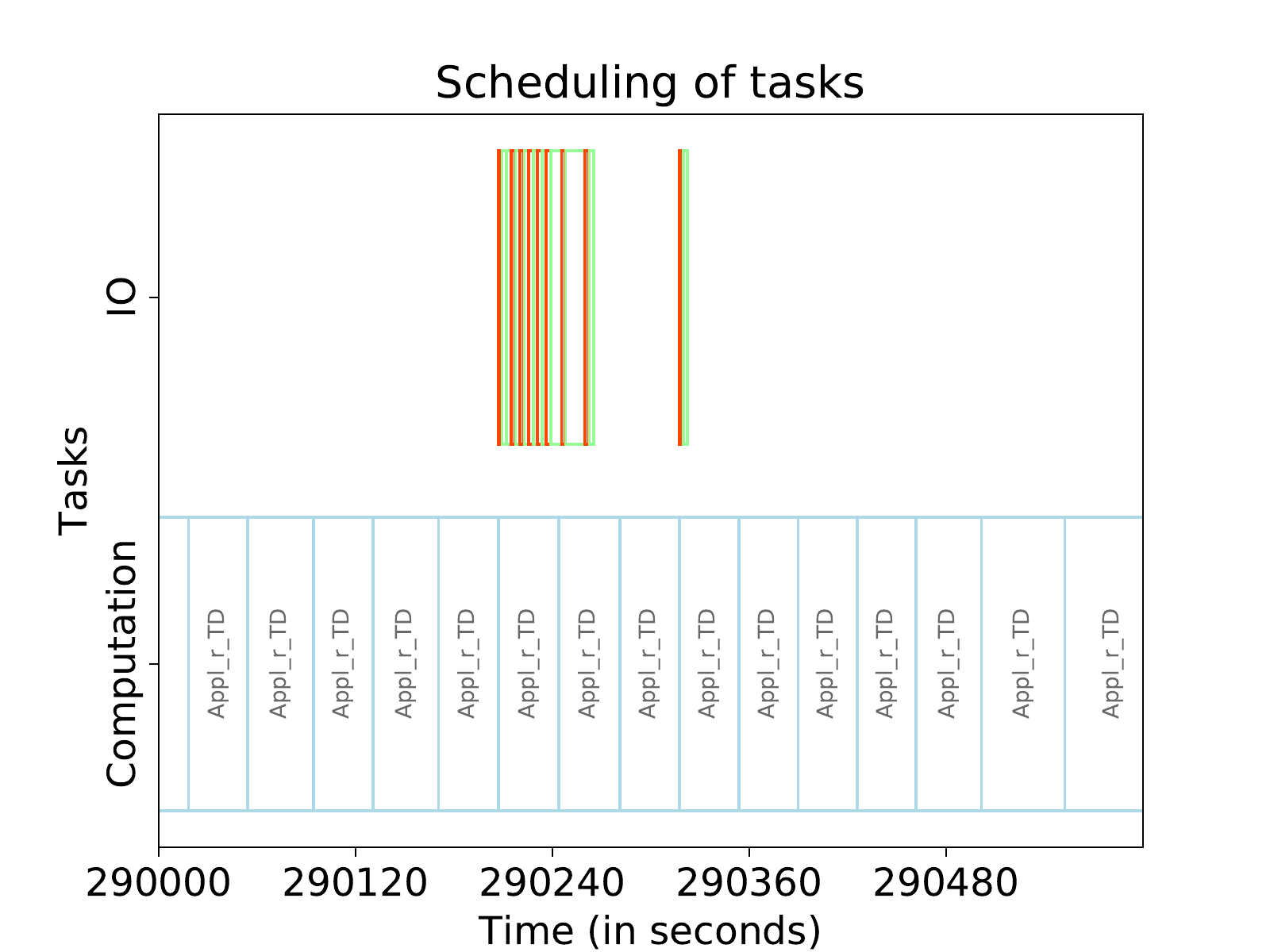} \\
\end{tabular}
\end{center}
\bfvspace
\caption{Example of the scheduling of the tasks in the execution
of the \randUTV{} factorization
of a $348{,}160 \times 348{,}160$ matrix in \texttt{ut}
in two different periods.
The left plot shows the scheduling during one hour;
the right plot is a zoom of the first ten minutes of the left plot.}
\label{fig:scheduling_ut}
\end{figure}

Figure~\ref{fig:scheduling_ut}
shows a real example of the scheduling of the different tasks during some time
of the \randUTV{} factorizations
of a $348{,}160 \times 348{,}160$ matrix in \texttt{ut}
when no orthonormal matrices are built and $q=0$.
The left plot shows the scheduling during one hour of the experiment;
the right plot is a zoom of the first ten minutes of the left plot.
The top part of the two plots
shows the I/O tasks performed by the application:
A red rectangle is a write operation, whereas
a green rectangle is a read operation.
The bottom part of the two plots
shows the computation performed by the application.
The names of the tasks are only shown when there is enough room.
As can be seen,
the overlapping of computation and I/O is almost perfect:
The computer performs I/O operations and computation at the same time.
Another interesting remark obtained from this plot
is the fact that the I/O operations are usually smaller than
the computational operations
despite the high processing power of the computer (36 cores),
which makes the application still computation-bound in this case (\texttt{ut}).
This means that higher computational power could further reduce
the processing time of the factorizations.

\subsubsection{Performances of Out-Of-Core codes versus In-Core codes}


\begin{figure}[ht!]
\tfvspace
\begin{center}
\begin{tabular}{cc}
\includegraphics[width=0.45\textwidth]{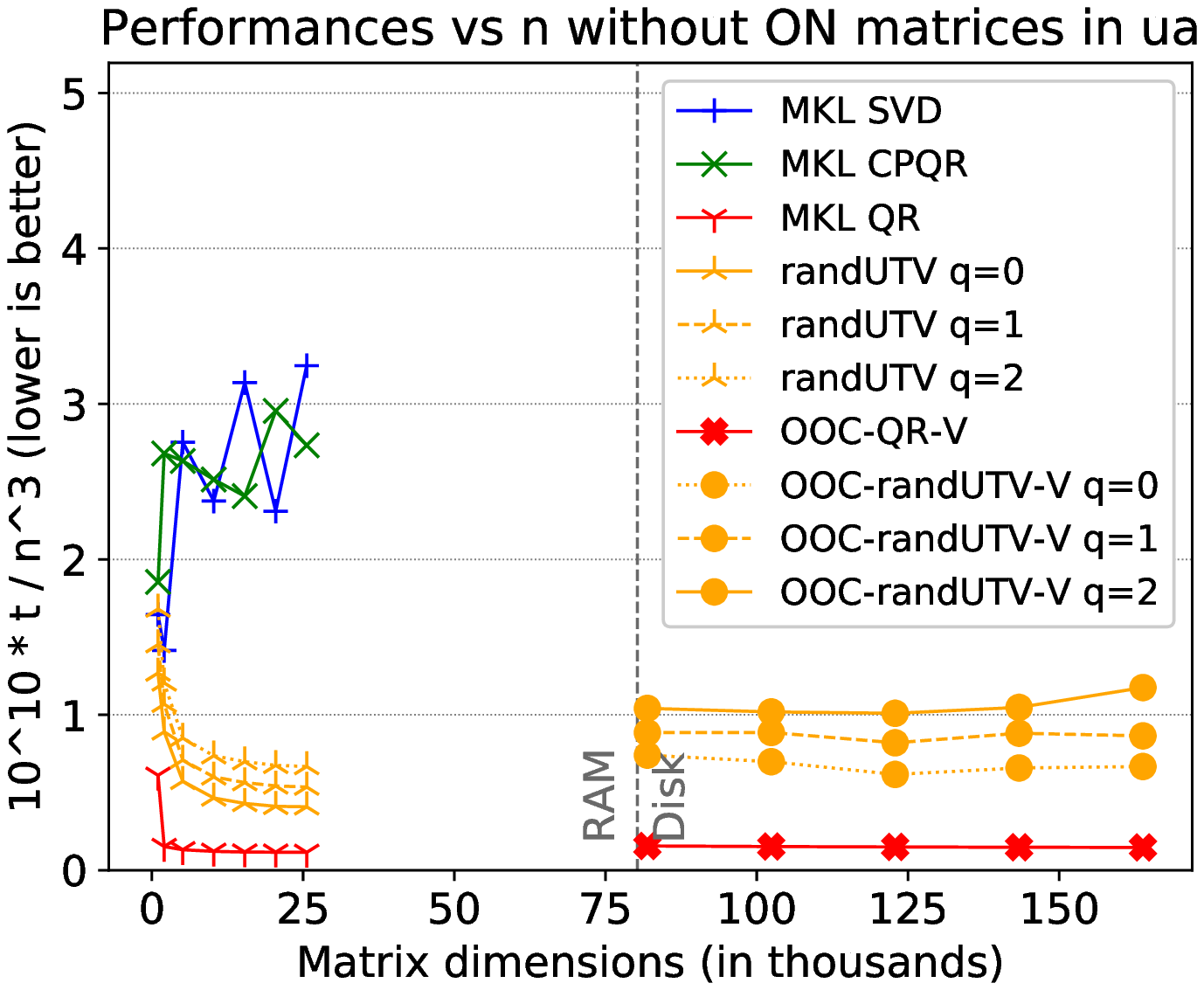} &
\includegraphics[width=0.45\textwidth]{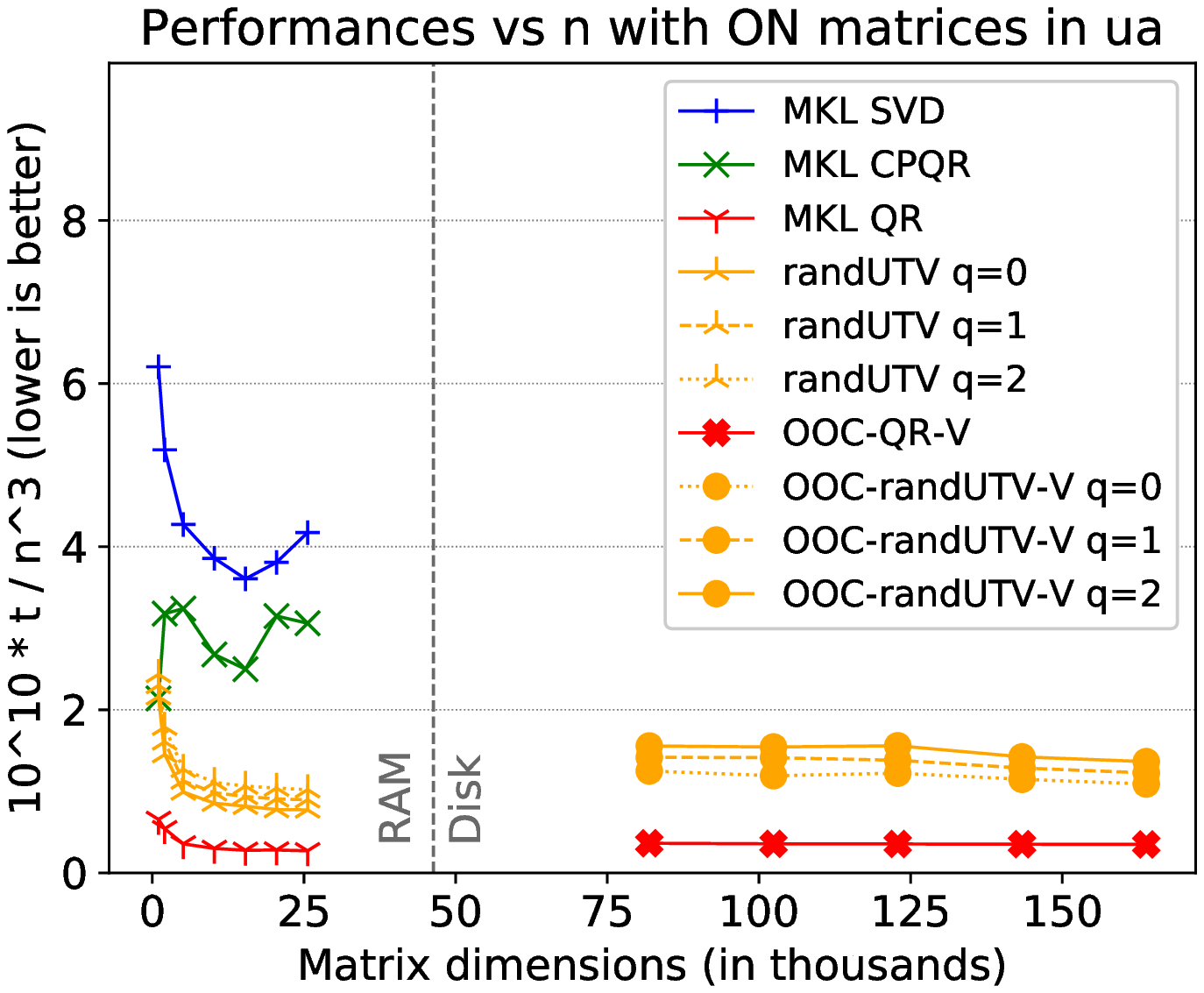} \\
\includegraphics[width=0.45\textwidth]{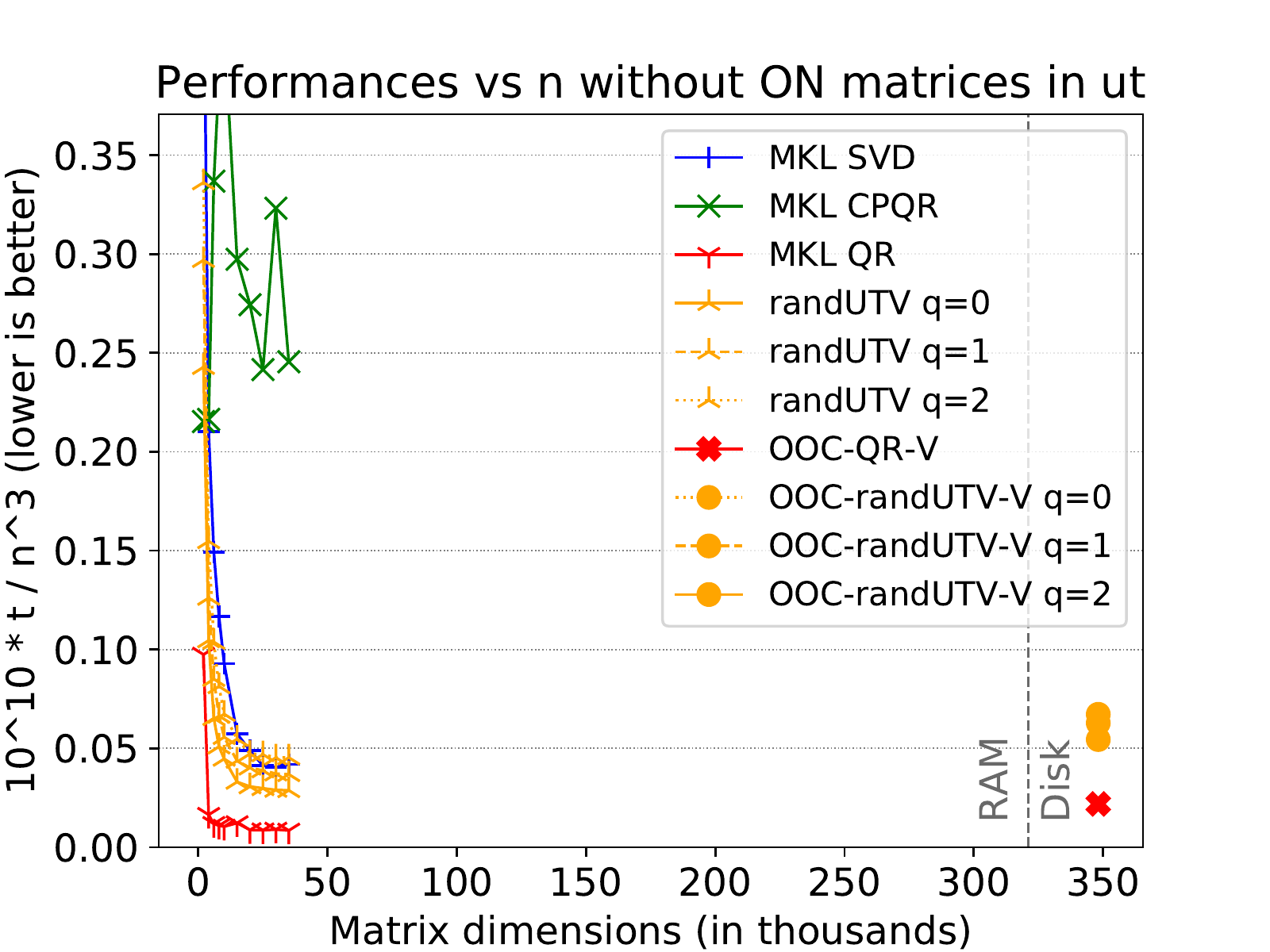} &
\includegraphics[width=0.45\textwidth]{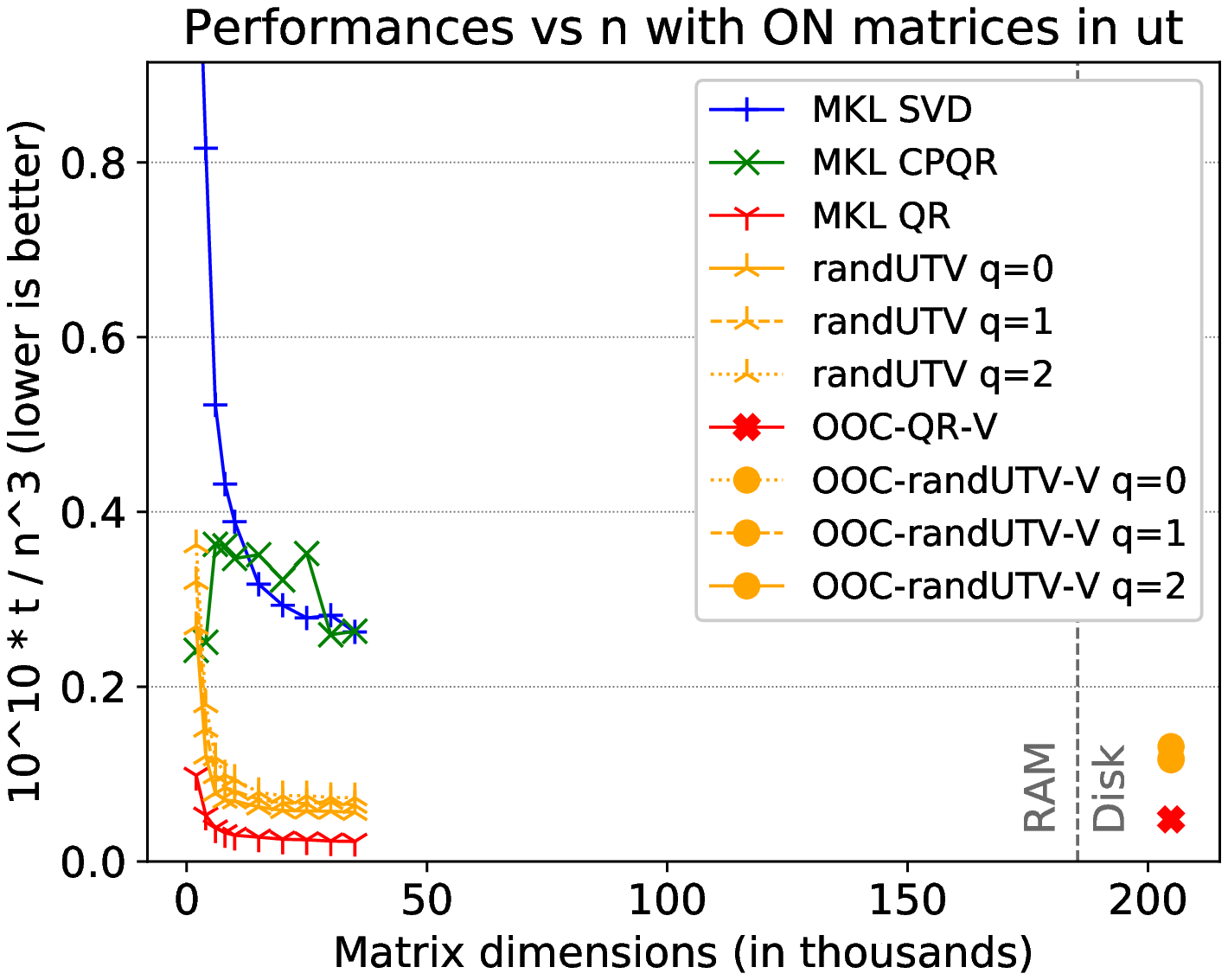} \\
\includegraphics[width=0.45\textwidth]{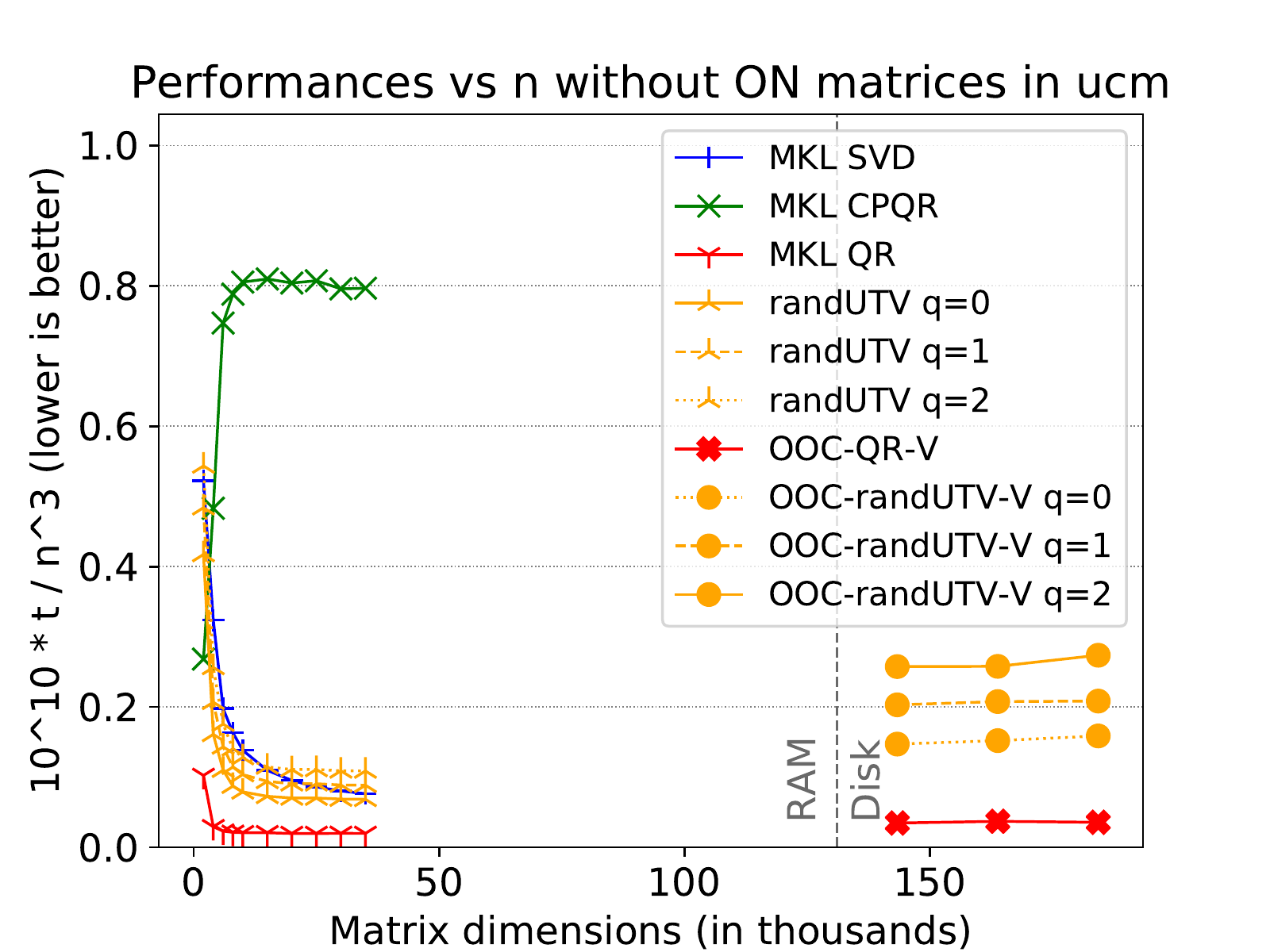} &
\includegraphics[width=0.45\textwidth]{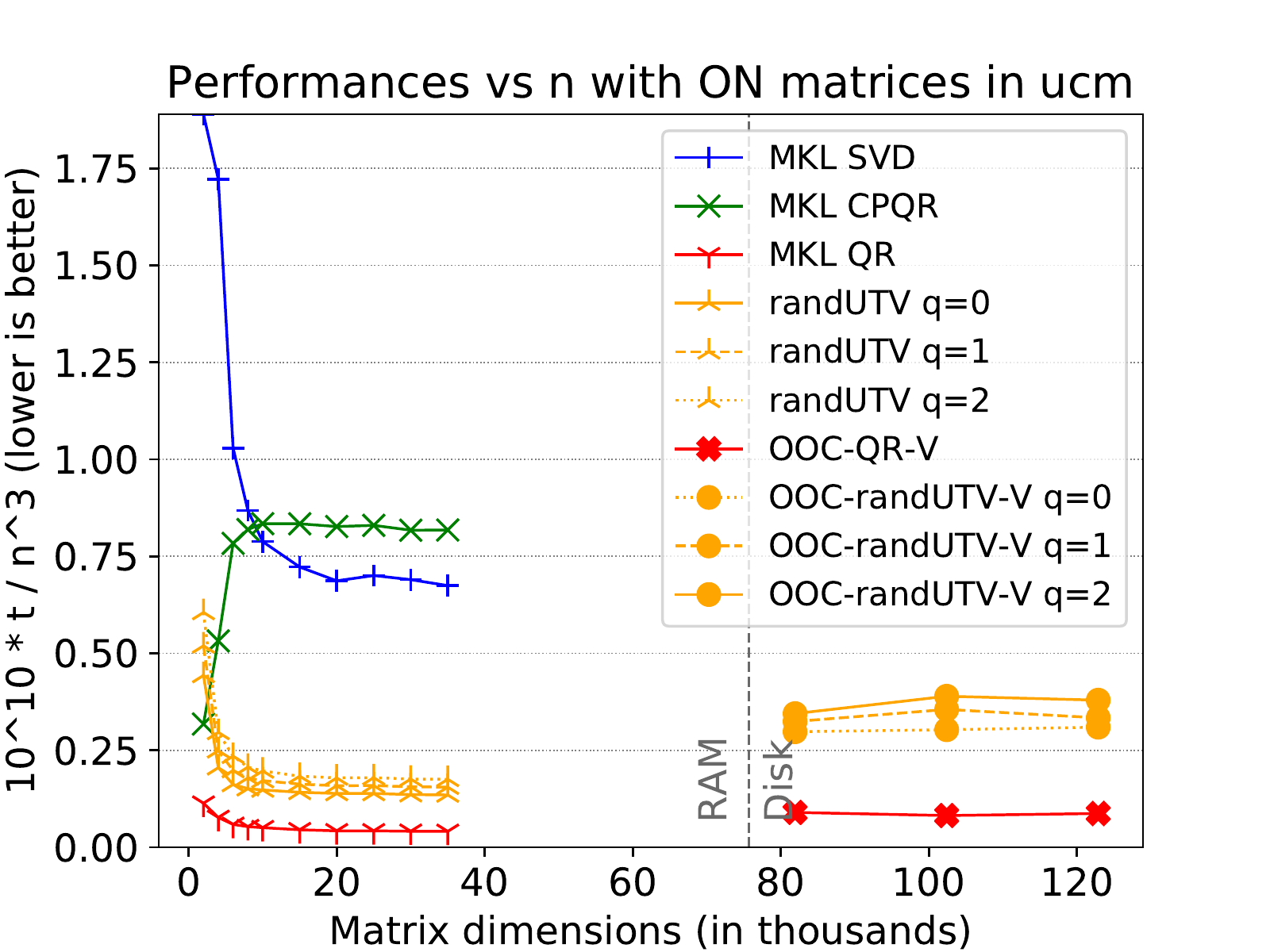} \\
\end{tabular}
\end{center}
\bfvspace
\caption{Performances versus matrix dimensions
for the best implementations of the QR and \randUTV{} factorizations.
In-core implementations are included as a reference.
Recall that the QR factorization does not reveal the rank,
and it is included as a reference.
The top row shows results in \texttt{ua},
the center row shows results in \texttt{ut}, and
the bottom row shows results in \texttt{ucm}.}
\label{fig:ooc_inc}
\end{figure}

Figure~\ref{fig:ooc_inc}
shows the performances of the best out-of-core implementations
as a function of the matrix dimensions.
This plot also shows performances of in-core factorizations
so that the speed of both types of implementations can be compared.
Notice that when no orthonormal matrices are built (left plots)
the in-core MKL SVD 
is much faster on \texttt{ut} and \texttt{ucm} than on \texttt{ua}.
In those cases, the MKL SVD is even much faster than the MKL CPQR
despite having a much higher computational cost.
These high performances are caused by the employment 
of a new implementation for computing the SVD in MKL 
that replaces the traditional implementation
for parallel architectures on matrices with dimensions larger than 4000.
We think that 
some of the techniques employed by Intel 
for this new implementation of the SVD in MKL 
could also be applied to our randUTV to make it faster when no orthornormal
matrices are built.

The top row of plots of Figure~\ref{fig:ooc_inc}
shows the results obtained in \texttt{ua}.
In these results
the best out-of-core QR factorization is
about 28 \% slower than the in-core QR factorization
when no orthonormal matrices are built,
and
about 29 \% slower than the in-core QR factorization
when orthonormal matrices are built.
On the other side,
the out-of-core \randUTV{} is about
51 \% ($q=0$), 53 \% ($q=1$), and 51 \% ($q=2$)
slower than the in-core \randUTV{} factorization
when no orthonormal matrices are built,
and about
41 \% ($q=0$), 36 \% ($q=1$), and 33 \% ($q=2$)
slower than the in-core \randUTV{} factorization
when orthonormal matrices are built.
In all these cases the comparison has been performed
considering the best performance of the out-of-core implementations
against
the best performance of the in-core factorizations
(which is usually obtained on very large matrices).

The center row of plots of Figure~\ref{fig:ooc_inc}
shows the results obtained in \texttt{ut}.
In these results
the out-of-core QR factorization is
about 2.58 times as slow as the in-core QR factorization
when no orthonormal matrices are built,
and
about 2.10 tiimes as slow as the in-core QR factorization
when orthonormal matrices are built.
On the other side,
the out-of-core \randUTV{} is
1.89 ($q=0$), 1.70 ($q=1$), and 1.48 ($q=2$)
times as slow as the in-core \randUTV{} factorization
when no orthonormal matrices are built,
and about
2.07 ($q=0$), 1.83 ($q=1$), and 1.80 ($q=2$)
times as slow as the in-core \randUTV{} factorization
when orthonormal matrices are built.

The bottom row of plots of Figure~\ref{fig:ooc_inc}
shows the results obtained in \texttt{ucm}.
In these results
the out-of-core QR factorization is
about 1.76 times as slow as the in-core QR factorization
when no orthonormal matrices are built,
and
about 2.00 times as slow as the in-core QR factorization
when orthonormal matrices are built.
On the other side,
the out-of-core \randUTV{} is about
2.15 ($q=0$), 2.29 ($q=1$), and 2.36 ($q=2$)
times as slow as the in-core \randUTV{} factorization
when no orthonormal matrices are built,
and about
2.10 ($q=0$), 2.04 ($q=1$), and 2.03 ($q=2$)
times as slow as the in-core \randUTV{} factorization
when orthonormal matrices are built.

Therefore,
as can be seen,
in the worst cases
the speed of revealing
the rank of matrices stored in the slow external devices
is only about two times as slow as
the speed of revealing
the rank of matrices already stored in the main memory.

In conclusion,
our out-of-core implementations of the \randUTV{} factorization
are able to efficiently process very large data that do not fit in RAM and
must be stored in the disk drive.
Despite the slow speed of the hard drive with respect to the main memory,
these methods are able to process matrices
with only a minor loss of performances.

\section{Conclusions}
\label{sec:conclusions}

This paper describes a set of algorithms for computing rank revealing
factorizations of matrices that are stored on external memory devices such as
solid-state or spinning disk hard drives. 
Standard techniques for computing rank revealing factorizations
of matrices perform very poorly in this environment, as they inherently consist
of a sequence of matrix-vector operations, which necessitate a large number of
read and write operations to the disk. We use randomization to reorganize the
computation to operate on large blocks of the matrix, thereby dramatically
reducing the amount of communication required.

Numerical experiments demonstrate that the speed of
the new randomized methods that operate on data stored
on external devices is comparable to traditional
methods that operate on data stored in main memory.
This enables the processing of very large matrices
in a cost efficient way. As the performance of solid
state hard drives is rapidly improving in terms of both
speed and capacity, the methods described are likely
to gain even more of a competitive advantage in coming years.

\section*{Acknowledgements}

P.G.~Martinsson acknowledges support from the Office of Naval Research
(award  N00014-18-1-2354), from the National Science Foundation (award
DMS-1620472), and from Nvidia Corp.
G. Quintana-Ort\'{\i} was supported by
the Spanish Ministry of Science, Innovation and Universities
under Grant RTI2018-098156-B-C54 co-financed with FEDER funds.
The authors would like to thank
Francisco D. Igual (Universidad Complutense de Madrid) and
Javier Navarrete (Universitat d'Alacant)
for granting access to the \texttt{ucm} and the \texttt{ua} servers,
respectively.


\bibliography{main_bib}
\bibliographystyle{amsplain}

\end{document}